\documentclass[prd,preprint,showpacs]{revtex4-1}
\usepackage[utf8]{inputenc}
\usepackage{amsmath}
\usepackage{latexsym}
\usepackage{amsfonts}
\usepackage{graphicx}
\usepackage{mathrsfs}
\usepackage{CJK}
\usepackage{longtable}

\usepackage{hyperref}
\hypersetup{
  colorlinks = true,
  urlcolor = blue,
  linkcolor = blue,
  citecolor = green,
  filecolor = magenta,
}

\newcommand{\ud}{\mathrm{d}}
\newcommand{\pd}{\partial}

\begin{document}
\begin{CJK*}{GB}{gbsn}

\title{Gravitational waves in Einstein-\ae ther and generalized TeVeS theory after GW170817}
\author{Yungui Gong}
\email{yggong@hust.edu.cn}
\affiliation{School of Physics, Huazhong University of Science and Technology, Wuhan, Hubei 430074, China}
\author{Shaoqi Hou}
\email{shou1397@hust.edu.cn}
\affiliation{School of Physics, Huazhong University of Science and Technology, Wuhan, Hubei 430074, China}
\author{Dicong Liang}
\email{dcliang@hust.edu.cn}
\affiliation{School of Physics, Huazhong University of Science and Technology, Wuhan, Hubei 430074, China}
\author{Eleftherios Papantonopoulos}
\email{lpapa@central.ntua.gr}
\affiliation{Department of Physics, National Technical University of Athens, Zografou Campus GR 157 73, Athens, Greece}

\begin{abstract}
In this work, we discuss the polarization contents of Einstein-\ae ther theory and the generalized tensor-vector-scalar (TeVeS) theory, as both theories have a normalized timelike vector field.
We derive the linearized equations of motion around the flat spacetime background using the gauge-invariant variables to easily separate physical degrees of freedom.
We find the plane wave solutions are then found, and identify the polarizations  by examining the geodesic deviation equations.
We find that there are five polarizations in Einstein-\ae ther theory and six polarizations
in the generalized TeVeS theory. In particular, the transverse breathing mode is mixed with the pure longitudinal mode.
We also discuss the experimental tests of the extra polarizations in Einstein-\ae ther theory  using pulsar timing arrays combined with the gravitational-wave speed bound derived from the observations on GW 170817 and GRB 170817A.
It turns out that it {might} be difficult to use pulsar timing arrays to distinguish different polarizations in Einstein-\ae ther theory.
The same speed bound also forces one of the propagating modes in the generalized TeVeS theory to travel much faster than the speed of light.
{Since the strong coupling problem does not exist in some parameter subspaces, the generalized TeVeS theory is excluded in these parameter subspaces.}
\end{abstract}


\maketitle
\end{CJK*}

\section{Introduction}

The direct detection of gravitational waves (GWs) by the LIGO Scientific and Virgo Collaborations marks the beginning of the era  of testing General Relativity (GR) in the strong-field regime \cite{Abbott:2016blz,Abbott:2016nmj,Abbott:2017vtc,Abbott:2017oio,TheLIGOScientific:2017qsa,Abbott:2017gyy}.
In particular, the detection of GW170814  confirmed the polarization content of GWs for the first time,
and the analysis showed that the pure tensor polarizations are
favored against pure vector and pure scalar polarizations \cite{Abbott:2017oio}.
GW170817 was the first event of a binary neutron star merger.
Together with its electromagnetic counterpart---the gamma-ray burst GRB 170817A  \cite{TheLIGOScientific:2017qsa,Goldstein:2017mmi,Savchenko:2017ffs}---they not only provided a very tight bound on the speed of GWs, but also heralded a new age of multimessenger astrophysics.
While ground-based interferometers detect GWs in the high-frequency band (10$-10^4$ Hz), pulsar timing arrays (PTAs) \cite{Kramer:2013kea,Hobbs:2009yy,McLaughlin:2013ira,Hobbs:2013aka} are sensitive to GWs in the lower-frequency band (around $10^{-10}$$-10^{-6}$ Hz) \cite{Moore:2014lga}.
The intermediate-frequency band can be best probed by eLISA \cite{Seoane:2013qna}, TianQin \cite{Luo:2015ght}, TaiJi,
the DECi-hertz Interferometer Gravitational wave Observatory  \cite{Kawamura:2011zz}
and the recently proposed Mid-band Atomic Gravitational Wave Interferometric Sensor (MAGIS) \cite{Graham:2017pmn}.
So PTAs, eLISA and MAGIS will provide  tests of GWs that are complementary to LIGO/Virgo.

In general, GWs have at most six polarizations \cite{Eardley:1974nw}.
Alternative theories of gravity to GR predict extra polarizations,
in addition to the familiar plus and cross polarizations in GR \cite{Will:2014kxa}.
These extra polarizations are usually excited by the extra d.o.f. contained in alternative theories of gravity.
For example, in scalar-tensor theories of gravity, the massless scalar field excites the transverse breathing polarization,
while the massive one  excites the longitudinal polarization \cite{Will:2014kxa,Liang:2017ahj,Hou:2017bqj,Gong:2017bru}.
More complicated alternative theories of gravity will add more polarizations, such as Einstein-\ae ther theory \cite{Jacobson:2000xp,Jacobson:2004ts} and the generalized tensor-vector-scalar (TeVeS) theory \cite{Seifert:2007fr,Sagi:2010ei},
whose GW polarization contents are the topics of the present work.
Both theories have the normalized timelike vector fields, which break the local Lorentz invariance (LLI).
We will develop a gauge-invariant formalism to calculate the polarizations
of GWs in modified gravitational theories like Einstein-\ae ther theory and the generalized TeVeS theory,
so that the physical d.o.f. are separated automatically, and GW solutions can be obtained in an arbitrary gauge.
We will also present bounds on the parameters respecting the recent observational results on GWs \cite{TheLIGOScientific:2017qsa,Goldstein:2017mmi,Savchenko:2017ffs,Monitor:2017mdv}.

Einstein-\ae ther theory  is a local Lorentz-violating theory of gravity \cite{Jacobson:2000xp}.
The gravitational interaction is mediated by the metric tensor $g_{\mu\nu}$ and a unit timelike vector field $u^\mu$.
Since $u^\mu$ never vanishes and pervades the Universe, it is called the ``\ae ther" field.
It breaks LLI, as it defines a preferred frame everywhere in the spacetime.
GW solutions have already been  obtained in Ref.~\cite{Jacobson:2004ts}
in the flat spacetime background, where the \ae ther field $u^\mu$ is at rest.
It was found out that there are generally three extra polarizations, excited by the three d.o.f. of the \ae ther field $u^\mu$.
Each polarization propagates at a speed different from 1 in a broad range of parameter space, although they are all massless.
In the present work, GW solutions will be derived again using the gauge-invariant variables.
The polarization contents of GWs are then discussed.
With the recent bound on GW speed inferred from the observations of GW170817 and GRB 170817A \cite{Monitor:2017mdv}, one sets bounds on the parameters in this theory,
and thus predicts the possibility of detecting polarizations with PTAs by calculating the cross-correlation functions for different polarizations.
The results show that the cross-correlation functions take very similar forms for different polarizations in some parameter regions,
so it will be difficult to use PTAs to distinguish polarizations, or to examine whether there are extra polarizations.
However, there exist other parameter regions, in which the cross-correlation functions vary a lot with different polarizations,
which makes it possible to use PTAs to distinguish polarizations. 
The authors of Ref.~\cite{Chesler:2017khz} excluded generalized Einstein-\ae ther theories \cite{Zlosnik:2006zu} based on GW150914 \cite{Abbott:2016blz}.

TeVeS theory, was originally proposed by Bekenstein to solve the dark matter problem \cite{Bekenstein:2004ne}.
It reduces to Milgrom's modified Newtonian dynamics (MOND) \cite{Milgrom:1983ca,Milgrom:1983pn,Milgrom:1983zz} in the nonrelativistic limit.
In this theory, there are three fields mediating gravity: the ``Einstein metric" tensor $g_{\mu\nu}$,
a unit timelike vector field $\mathfrak U^\mu$, and a scalar field $\sigma$.
Matter fields minimally couple to the physical metric which is related to the Einstein
metric via the disformal transformation $\tilde g_{\mu\nu}=e^{-2\sigma}g_{\mu\nu}-2\mathfrak U_\mu \mathfrak U_\nu \sinh(2\sigma)$.
The action of $\mathfrak U^\mu$ is of the Maxwellian type, a special form included in the \ae ther's action.
However, TeVeS theory suffers from some problems such as  instability in the spherically symmetric solutions, and these problems could be cured by allowing the action of $\mathfrak U^\mu$ to be the most general one, i.e., that of the \ae ther field \cite{Seifert:2007fr}.
The theory thus obtained  is called the generalized TeVeS theory.
Sagi has already discussed the GW solutions in the generalized TeVeS theory and its polarization contents \cite{Sagi:2010ei}.
In the present work, the GW polarization content will be briefly analyzed again in a gauge-invariant way.
We will also discuss the implications of the bound on the speed of GWs in this theory.
The cosmological constraints on these alternative theories were discussed in Refs. \cite{Baker:2017hug,Zlosnik:2006zu,Pasqua:2015rxa}.

This work is organized in the following way.
First, in Sec.~\ref{sec-einae-gw} we discuss the GW solutions around the flat spacetime background in Einstein-\ae ther theory.
In particular, after a brief introduction to Einstein-\ae ther theory,
we solve the equations of motion using the gauge-invariant variables in Sec.~\ref{sec-eom-ginv},
and the polarization content of GWs is thus obtained in Sec.~\ref{sec-einae-pols}.
We discuss the experimental constraints on Einstein-\ae ther theory in Sec.~\ref{sec-einae-cons}.
In Sec.~\ref{sec-einae-ptas}, we compute the cross-correlation functions
for different polarizations by taking into account the speed bound on GW propagation.
Second, we discuss the GW solutions and the polarization content of the generalized TeVeS theory  in Sec. \ref{sec-gtvs}.
Again, after a brief introduction, we obtain
the GW solutions (mainly for the scalar field $\sigma$) and analyze the polarization content  in Sec.~\ref{sec-gtvs-gw}.
In Sec.~\ref{sec-teves-cons} we discuss the constraints on the generalized TeVeS theory.
Finally, in Sec. \ref{sec-con} we summarize our work.
Throughout this work, we use units such that the speed of light in vacuum is $c=1$.

\section{Gravitational Waves in Einstein-\AE ther Theory}\label{sec-einae-gw}

The action of Einstein-\ae ther theory is given by \cite{Jacobson:2004ts}
\begin{equation}\label{aeact}
\begin{split}
  S_{\text{EH-\ae}} =&\frac{1}{16\pi G} \int\ud^4x\sqrt{-g}[R-c_1(\nabla_\mu u_\nu)\nabla^\mu u^\nu-c_2(\nabla_\mu u^\mu)^2-c_3(\nabla_\mu u_\nu)\nabla^\nu u^\mu\\
  &+c_4(u^\rho\nabla_\rho u^\mu)u^\sigma\nabla_\sigma u_\mu+\lambda(u^\mu u_\mu+1)],
  \end{split}
\end{equation}
where $\lambda$ is a Lagrange multiplier, $G$ is the gravitational coupling constant, and
the constants $c_i\,(i=1,2,3,4)$ are expected to be of the order unity.
The Lagrange multiplier $\lambda$ renders $u^\mu$ a normalized timelike vector field, which defines a preferred reference frame at each spacetime point.
LLI is thus violated.
Let $S_m[g_{\mu\nu},\psi_m]$ be the matter action where $\psi_m$ collectively represents the matter fields.
The field $\psi_m$ is assumed to minimally couple with $g_{\mu\nu}$, so test particles follow geodesics in free fall.
In the following section, the GW solutions will be obtained by expressing the linearized equations of motion in terms of the gauge-invariant variables.

\subsection{Equations of motion}\label{sec-eom-ginv}

Ignoring the matter sector of the action, the equations of motion are obtained with the variational principle given below:
\begin{gather}
  R_{\mu\nu}-\frac{1}{2}g_{\mu\nu}R= T_{\mu\nu}^{\ae},\label{eq-ae-1}\\
  c_1\nabla_\mu\nabla^\mu u_\nu+c_2\nabla_\nu\nabla_\mu u^\mu+c_3\nabla_\mu\nabla_\nu u^\mu
  \nonumber\\
   -c_4  \nabla_\mu(u^\mu a_\nu)+c_4a_\mu\nabla_\nu u^\mu+\lambda u_\nu= 0,\label{eq-ae-2}\\
   u^\mu u_\mu+1=0,\label{eq-ae-3}
\end{gather}
where $a^\mu=u^\nu\nabla_\nu u^\mu$ is the 4-acceleration of $u^\mu$ and the \ae ther stress-energy tensor $T_{\mu\nu}^{\ae}$ is
\begin{eqnarray}
  T_{\mu\nu}^{\ae} &=& \lambda[u_\mu u_\nu-\frac{1}{2}g_{\mu\nu}(u^\rho u_\rho+1)]+c_1[(\nabla_\mu u_\rho)\nabla_\nu u^\rho-(\nabla_\rho u_\mu)\nabla^\rho u_\nu+\nabla_\rho(u_{(\mu}\nabla^\rho u_{\nu)}
  \nonumber\\
  &&-u_{(\mu}\nabla_{\nu)}u^\rho+u^\rho\nabla_{(\mu}u_{\nu)})]+c_2g_{\mu\nu}\nabla_\rho(u^\rho\nabla_\sigma u^\sigma)+c_3\nabla_\rho(u_{(\mu}\nabla_{\nu)}u^\rho-u_{(\mu}\nabla^\rho u_{\nu)}
  \nonumber\\
  &&+u^\rho\nabla_{(\mu}u_{\nu)})+c_4[a_\mu a_\nu-\nabla_\rho(2u^\rho u_{(\mu}a_{\nu)}-a^\rho u_\mu u_\nu)]
  \nonumber\\
  &&+\frac{1}{2}g_{\mu\nu}[-c_1(\nabla_\rho u_\sigma)\nabla^\rho u^\sigma-c_2(\nabla_\rho u^\rho)^2-c_3(\nabla_\rho u_\sigma)\nabla^\sigma u^\rho+c_4a_\rho a^\rho].
\end{eqnarray}
Here, Eq.~\eqref{eq-ae-3} is a constraint equation.

In the following, we will look for GW solutions  around the flat spacetime background, with
the zeroth-order solution given by
\begin{equation}\label{0thsol}
  g_{\mu\nu}=\eta_{\mu\nu},\quad u^\mu=\underline u^\mu=(1,0,0,0).
\end{equation}
Now, we perturb the metric and the \ae ther field in the following way:
\begin{equation}\label{eq-pbs}
  g_{\mu\nu}=\eta_{\mu\nu}+h_{\mu\nu},\quad u^\mu=\underline{u}^\mu+v^\mu.
\end{equation}
We decompose the metric perturbation $h_{\mu\nu}$ and the perturbed \ae ther field $v^\mu$ in the following way \cite{Flanagan:2005yc}:
\begin{gather}
  h_{tt} = 2\phi, \label{httdec}\\
  h_{tj} = \beta_j+\partial_j\gamma, \label{htjdec}\\
  h_{jk} = h_{jk}^\mathrm{TT}+\frac{1}{3}H\delta_{jk}+\partial_{(j}\epsilon_{k)}+\left(\partial_j\partial_k-\frac{1}{3}\delta_{jk}\nabla^2\right)\rho,\label{hjkdec}\\
  v^0=\frac{1}{2}h_{00}=\phi,\label{v0dec}\\
   v^j=\mu^j+\partial^j\omega.\label{vjdec}
\end{gather}
In the above expressions, $h_{jk}^\mathrm{TT}$ is the transverse-traceless part of $h_{jk}$, satisfying $\pd^kh_{jk}^\mathrm{TT}=0$ and $\eta^{jk}h_{jk}^\mathrm{TT}=0$. $\beta_j,\,\epsilon_j$ and $\mu^j$ are transverse vectors.
Equation~\eqref{v0dec} is the consequence of $u^\mu u_\mu=-1$.
Under the infinitesimal coordinate transformation $x^\mu\rightarrow x^\mu+\xi^\mu$, one has
\begin{gather}\label{eq-gt}
  h_{\mu\nu}\rightarrow h_{\mu\nu}-\partial_\mu\xi_\nu-\partial_\nu\xi_\mu,\\
  u^\mu\rightarrow u^\mu+\underline u^\nu\partial_\nu\xi^\mu.
\end{gather}
If an infinitesimal coordinate transformation is generated by $\xi_\mu=(\xi_t,\xi_j)=(A,B_j+\partial_jC)$ with $\partial^jB_j=0$, it can be shown that \cite{Flanagan:2005yc}
\begin{gather}
  \phi\rightarrow\phi-\dot A,\;\beta_j\rightarrow\beta_j-\dot B_j,\;\gamma\rightarrow\gamma-A-\dot C, \\
  H\rightarrow H-2\nabla^2C,\;\rho\rightarrow\rho-2C,\;\epsilon_j\rightarrow\epsilon_j-2B_j,\\
  h_{jk}^\mathrm{TT}\rightarrow h_{jk}^\mathrm{TT},
\end{gather}
where a dot denotes a partial time derivative and $\nabla^2=\partial_j\partial^j$ is the Laplacian.
The gauge transformation of the \ae ther field is
\begin{equation}\label{gtae}
  \mu^j\rightarrow\mu^j+\dot B^j,\quad\omega\rightarrow\omega+\dot C.
\end{equation}
Therefore, gauge-invariant variables can be defined \cite{Flanagan:2005yc}, which are $h_{jk}^\mathrm{TT}$ and
\begin{gather}
  \Phi = -\phi+\dot\gamma-\frac{1}{2}\ddot\rho, \\
  \Theta = \frac{1}{3}(H-\nabla^2\rho), \\
  \Xi_j = \beta_j-\frac{1}{2}\dot\epsilon_j,\\
  \Sigma_j=\beta_j+\mu_j,\\
   \Omega=\omega+\frac{1}{2}\dot\rho.
\end{gather}
There are in total nine gauge-invariant variables.
This is expected, as of the originally fourteen variables
the general covariance of the action \eqref{aeact} removes four d.o.f., and the constraint \eqref{eq-ae-3} removes one more.
The equations of motion \eqref{eq-ae-1} and \eqref{eq-ae-2} will remove four more d.o.f., leaving five physical d.o.f..

After some straightforward but tedious algebraic manipulations, we get
\begin{gather}
  \frac{c_{14}}{2-c_{14}}[c_{123}(1+c_2+c_{123})-2(1+c_2)^2]\ddot{\Omega}+c_{123}\nabla^2\Omega=0, \label{eomscl}\\
  c_{14}\ddot\Sigma_j-\frac{c_1-c_1^2/2+c_3^2/2}{1-c_{13}}\nabla^2\Sigma_j=0,\label{eomsig} \\
  \frac{1}{2}(c_{13}-1)\ddot h_{jk}^\mathrm{TT}+\frac{1}{2}\nabla^2h_{jk}^\mathrm{TT}=0,\label{ettjk}
\end{gather}
where $c_{13}=c_1+c_3$, $c_{14}=c_1+c_4$, and $c_{123}=c_1+c_2+c_3$.
So there are only five propagating physical d.o.f..
Two of them are tensor d.o.f. represented by $h_{jk}^\mathrm{TT}$, another two are vector
d.o.f. given by $\Sigma_j$, and the remaining one is a scalar d.o.f. given by $\Omega$.
The squared speeds of these modes can be easily read off from the above equations, and they are
\begin{gather}
  s_g^2=\frac{1}{1-c_{13}},\label{tenspd}\\
  s_v^2=\frac{c_1-c_1^2/2+c_3^2/2}{c_{14}(1-c_{13})},\label{vecspd}\\
  s_s^2=\frac{c_{123}(2-c_{14})}{c_{14}(1-c_{13})(2+2c_2+c_{123})},\label{sclspd}
\end{gather}
respectively.
These speeds are generally different from one another and from 1.
When $c_{13}=c_4=0$ and $2c_1c_2=c_2-c_1$ are satisfied, they are simultaneously one.
The remaining gauge-invariant variables are given by
\begin{gather}
  \Phi=\frac{c_{14}-2c_{13}}{2-c_{14}}\dot\Omega,\label{phom} \\
  \Theta=\frac{2c_{14}(c_{13}-1)}{2-c_{14}}\dot\Omega,\label{thom}\\
  \Xi_j=-\frac{c_{13}}{1-c_{13}}\Sigma_j.\label{xisi}
\end{gather}
These are dependent variables.
In deriving these relations, one imposes  the following conditions
\begin{equation}\label{eq-conds}
  c_{13}\ne1,\quad c_{14}\ne0,\quad c_{14}\ne2,\quad 3c_2\ne-2-c_{13}.
\end{equation}

\subsection{Polarizations of gravitational waves}\label{sec-einae-pols}

Since the matter fields are assumed to minimally couple with the metric tensor only,
the polarization content of GWs in Einstein-\ae ther theory is determined by examining the linearized geodesic deviation equation
\begin{equation}\label{eq-dev-l}
   \ddot x^j=\frac{d^2x^j}{dt^2}=-R_{tjtk}x^k,
\end{equation}
which describes the relative acceleration between two nearby test particles separated by the deviation vector $x^j$.
In terms of gauge-invariant variables, the electric components $R_{tjtk}$ of the Riemann tensor are given by  \cite{Flanagan:2005yc}
\begin{equation}\label{eq-rtitkginv}
  R_{tjtk}=-\frac{1}{2}\ddot h_{jk}^\mathrm{TT}+\dot\Xi_{(j,k)}+\Phi_{,jk}-\frac{1}{2}\ddot\Theta\delta_{jk}.
\end{equation}
To be more specific and due to the rotational symmetry of the Minkowski spacetime, one considers a situation where the plane GWs propagate in the $+z$ direction.
The wave vectors of the scalar, vector, and tensor modes are
\begin{gather}
  k_s^\mu=\omega_s(1,0,0,1/s_s),\label{aewv-s}\\
   k_v^\mu=\omega_v(1,0,0,1/s_v),\label{aewv-v}\\
    k_g^\mu=\omega_g(1,0,0,1/s_g),\label{aewv-g}
\end{gather}
respectively, where the $\omega$'s are the corresponding angular frequencies.
In this case, the nonvanishing components of $h_{jk}^\text{TT}$ are $h_{11}^\text{TT}=-h_{22}^\text{TT}=h_+$ and $h_{12}^\text{TT}=h_{21}^\text{TT}=h_\times$.
For the vector mode, $\Sigma_3=0$ since $\partial_j\Sigma^j=0$.

By calculating $R_{tjtk}$ we find that there are five polarization states.
In terms of $R_{tjtk}$, the plus polarization is given by $\hat P_+=-R_{txtx}+R_{tyty}=\ddot h_+$, and the cross polarization is $\hat P_\times=R_{txty}=-\ddot h_\times$;
the vector-$x$ polarization is represented by $\hat P_{xz}=R_{txtz}=-c_{13}\partial_3\dot\Sigma_1/[2(1-c_{13})]$,
and the vector-$y$ polarization is $\hat P_{yz}=R_{txty}=-c_{13}\partial_3\dot\Sigma_2/[2(1-c_{13})]$;
the transverse breathing polarization is specified by
$\hat P_b=R_{txtx}+R_{tyty}=-2c_{14}(c_{13}-1)\dddot\Omega/(2-c_{14})$,
and the longitudinal polarization is
$$\hat P_l=R_{tztz}=\frac{c_{14}-2c_{13}}{2-c_{14}}\partial_3^2\dot\Omega-
\frac{c_{14}(c_{13}-1)}{2-c_{14}}\dddot\Omega=\left[\frac{c_{14}-2c_{13}}{2-c_{14}}\frac{1}{s_s^2}-
\frac{c_{14}(c_{13}-1)}{2-c_{14}}\right]\dddot\Omega.$$
Among these polarizations, both the transverse breathing and the longitudinal modes are excited by the scalar d.o.f. $\Omega$,
so $\Omega$ excites a mixed state of $\hat P_b$ and $\hat P_l$, as in the case of Horndeski theory \cite{Hou:2017bqj,Gong:2017bru}.
One can also calculate the Newman-Penrose variables \cite{Newman:1961qr,Eardley:1973br,Eardley:1974nw},
and it is found that none of them vanish in general.

In the following discussion,  the gauge will be fixed so that
\begin{equation}\label{jmgauge}
  h_{0j}=0,\quad v^j{}_{,j}=0,
\end{equation}
which implies  that $\Sigma_j=\mu_j=v_j$ and $\dot\Omega=\frac{2-c_{14}}{2(c_{13}-1)}\phi$.
Therefore, one obtains
\begin{gather}
  h_+=e_+\cos[\omega_g(t-z/s_g)],\label{gw-plus} \\
  h_\times=e_\times\cos[\omega_g(t-z/s_g)], \label{gw-cross}\\
  v_j= \mu_j^0\cos[\omega_v(t-z/s_v)],\quad j=1,2,\label{gw-vec}\\
  \phi=\varphi\cos[\omega_g(t-z/s_s)],\label{gw-scl}
\end{gather}
where $e_+,\,e_\times,\,\mu_j^0$ and $\varphi$ are the amplitudes.

\subsection{Discussion on the constraints}\label{sec-einae-cons}

As mentioned before, LLI is violated.
This can be seen in the post-Newtonian formalism developed by Foster and Jacobson \cite{Foster:2005dk}.
The post-Newtonian parameters $\alpha_1$ and $\alpha_2$ are given by
\begin{gather}\label{gials}
\alpha_1=-\frac{8(c_3^2+c_1c_4)}{2c_1-c_1^2+c_3^2},\\
\alpha_2=\frac{(2c_{13}-c_{14})^2}{c_{123}(2-c_{14})}-\frac{12c_3c_{13}+2c_1c_{14}(1-2c_{14})+(c_1^2-c_3^2)(4-6c_{13}+7c_{14})}{(2-c_{14})(2c_1-c_1^2+c_3^2)}.
\end{gather}
These parameters together with $\alpha_3$ (which vanishes in Einstein-\ae ther theory) measure the preferred-frame effects at the post-Newtonian order \cite{Poisson2014}.
According to Ref.~\cite{Will:2014kxa}, $|\alpha_1|\lesssim10^{-4}$ from  the Lunar Laser Ranging experiments, and $|\alpha_1|\lesssim4\times10^{-5}$ based on the observation of PSR J1738+0333 \cite{Freire:2012mg}.
In addition, $|\alpha_2|\lesssim2\times10^{-9}$ was obtained using the observations of the millisecond pulsars B1937+21 and J1744-1134 \cite{Shao:2013wga,Shapiro:1999fn}.

Moreover, Newton's constant is found to be \cite{Carroll:2004ai,Foster:2005dk}
\begin{equation}\label{eq-nc-eae}
  G_\text{N}=\frac{G}{1-c_{14}/2},
\end{equation}
and the gravitational constant appearing in the Friedman equation is \cite{Carroll:2004ai}
\begin{equation}\label{eq-cc-eae}
  G_\text{cosmo}=\frac{G}{1+(c_{13}+3c_2)/2}.
\end{equation}
In contrast to GR, these two constants are not the same, so the expansion rate of the Universe is different from that predicted by GR even if the matter content is the same in the two theories.
Thus the ratio of the two constants should be constrained, for example, by the observed primordial ${}^4$He abundance \cite{Carroll:2004ai}
\begin{equation}\label{eq-rgg}
  \left|\frac{G_\text{cosmo}}{G_\text{N}}-1\right|<\frac{1}{8}.
\end{equation}

The energy carried away by the gravitational waves should be positive, which leads to the following conditions \cite{Jacobson:2008aj}:
\begin{gather}
\frac{2c_1-c_1^2+c_3^2}{1-c_{13}}>0,\label{eq-pe1}\\
c_{14}(2-c_{14})>0.\label{eq-pe2}
\end{gather}
Finally, all of the speeds \eqref{tenspd}--\eqref{sclspd} should be greater than 1 so that there is no gravitational Cherenkov radiation \cite{Elliott:2005va}.

The recent observation of GW170817 \cite{TheLIGOScientific:2017qsa} determined that  photons arrived at the Earth about 1.7 s later than the GWs,
which has been used to set bounds on GWs' speed \cite{Monitor:2017mdv},
\begin{equation}\label{splgw170817}
 -3\times10^{-15}\le\frac{v_\text{GW}-v_\text{EM}}{v_\text{EM}}\le7\times10^{-16},
\end{equation}
where $v_\text{GW}$ and $v_\text{EM}$ are the speeds of the GW and the photon, respectively.
Suppose the photon speed $v_\text{EM}$  is 1; then, the GW speed is bounded from above, i.e., $v_\text{GW}\le1+7\times10^{-16}$.
If the detected GW signal is a tensor wave, then one obtains
\begin{equation}\label{eq-c13b-gw}
  c_{13}\le1.4\times10^{-15}
\end{equation}
using the speed squared for the spin-2 graviton
$s_g^2=1/(1-c_{13})$.

Combining all of the constraints listed above, one can set bounds on the $c_i$'s.
Because $\alpha_1$ and $\alpha_2$ are constrained to be small by observations,
one can expand the theory in powers of $\alpha_1$ and $\alpha_2$ \cite{Yagi:2013qpa,Yagi:2013ava}.
At the leading order
\begin{gather}\label{eq-c2c4}
  c_2=\frac{c_{13}(c_3-2c_1)}{3c_1}, \\
  \quad c_4=-\frac{c_3^2}{c_1},
\end{gather}
by setting $\alpha_1$ and $\alpha_2$ to zero.
Although at this order the $\alpha$'s all vanish, the preferred-frame effects will show up at higher orders in $\alpha_1$ and $\alpha_2$.
Even if the $\alpha$'s vanish identically, LLI is still violated, as the $\alpha$'s only parametrize the violation of LLI at the post-Newtonian order.
Now, the parameter space reduces to two dimensions, and it is parametrized by $c_\pm=c_1\pm c_3$ with $c_+=c_{13}$.
The parameters $c_\pm$ are constrained by the requirements that the perturbation around the flat spacetime
background is stable and has positive energy \cite{Jacobson:2004ts},
and that there is no gravitational Cherenkov radiation \cite{Elliott:2005va}.
These lead to
\begin{equation}\label{eq-con-cpm}
  0\le c_+\le1,\quad 0\le c_-\le\frac{c_+}{3(1-c_+)}
\end{equation}
to the leading order in $\alpha_1$ and $\alpha_2$.
These constraints lead to the superluminal propagation of GWs in the flat spacetime background \cite{Jacobson:2004ts}.

Yagi \textit{et~al.} \cite{Yagi:2013qpa,Yagi:2013ava} put further constraints on $c_\pm$ from binary pulsar observations.
Together with the stability and no-Cherenkov-radiation requirements,
the binary pulsar observations have pushed the available parameter space ($c_+,c_-$) to a small corner, as shown in Fig.~1 in Ref.~\cite{Yagi:2013ava}.
Let $c_+$ saturate the bound \eqref{eq-c13b-gw}, i.e., $c_+=1.4\times10^{-15}$, so $s_g=1+7\times10^{-16}$.
A careful examination of Fig.~1 in Ref.~\cite{Yagi:2013ava} shows that $c_-\lesssim 0.32c_+$ and $c_+\lesssim 0.005$.
For  future computations we choose the  parametrization
\begin{equation}\label{eq-cminus}
  c_-=r_{-}c_+
\end{equation}
near $c_+=1.4\times10^{-15}$  with $r_-\lesssim 0.32$.
Then, by using the speeds of the vector and  scalar GWs discussed in the previous subsection, we obtain
\begin{gather}
s_v=\frac{1}{2}\sqrt{\frac{(1+r_-)(1+r_{-}-r_{-}c_+)}{r_{-}}}s_g,\\
s_s=\frac{s_g}{\sqrt{3r_{-}}}.
\end{gather}
If $r_{-}=0.1,0.2$, or $0.3$, one gets three sets of speeds, which are listed in Table~\ref{tab-spds}.
As it shows, all speeds exceed 1 and decrease with $r_{-}$.
\begin{table}
  \centering
    \caption{The speeds of the vector and scalar GWs.}\label{tab-spds}
  \begin{tabular}{cccc}
    \hline\hline
    $r_{-}$ & 0.1 & 0.2 & 0.3 \\
    \hline
    $s_v$ & 1.74 & 1.34 & 1.19 \\
    $s_s$ & 1.83 & 1.29 & 1.05 \\
    \hline
  \end{tabular}
\end{table}
One can also check that with the chosen $r_-$, all $c_i$'s are of the order of $10^{-15}$.
The smallness of these parameters requires severe fine-tuning.

\begin{figure}
  \centering
  \includegraphics[width=0.5\textwidth]{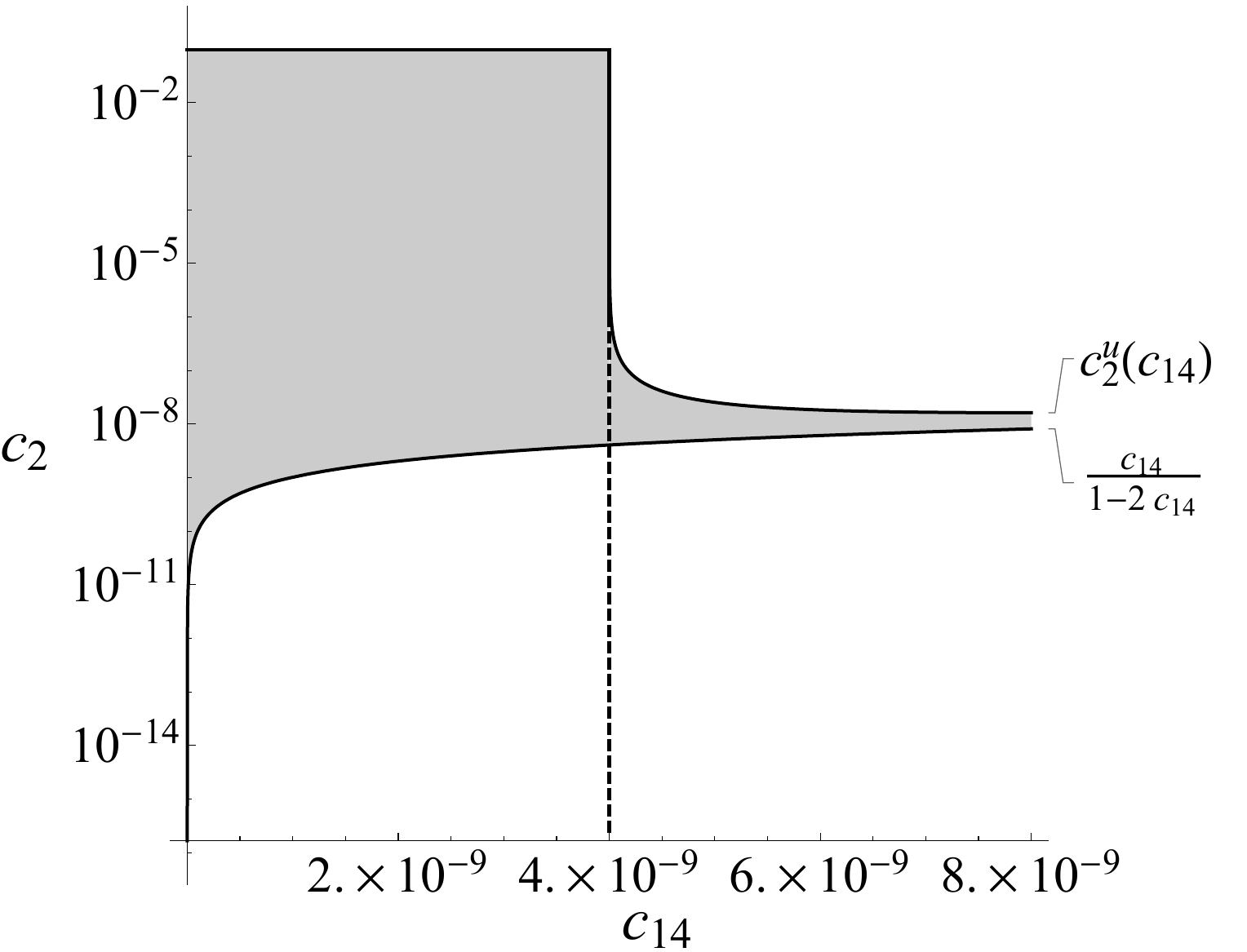}
  \caption{The constraints on $c_2$ in the range $0<c_{14}<8\times10^{-9}$.
  The shaded region is allowed.
  As $c_{14}$ increases, the upper and the lower bounds approach each other.
  Note that the vertical axis  uses a logarithmic scale.}\label{fig-c2}
\end{figure}

One may also let $c_{13}=0$ without setting $\alpha_1=\alpha_2=0$ as done in Ref.~\cite{Oost:2018tcv}.
In this case, $s_g=1$, i.e., the tensor GW propagates at the exact speed of light, and
\begin{equation}\label{eq-spds}
  s_v^2=\frac{c_1}{c_{14}},\quad s_s^2=\frac{c_2(2-c_{14})}{c_{14}(2+3c_2)}.
\end{equation}
In addition, $\alpha_1$ and $\alpha_2$ reduce to
\begin{equation}\label{eq-alphas}
  \alpha_1=-4c_{14},\quad\alpha_2=\frac{c_{14}[c_2-c_{14}(1+2c_2)]}{c_2(c_{14}-2)}.
\end{equation}
Using the constraints on $s_v,s_s,\alpha_1$, and $\alpha_2$ together with the inequalities \eqref{eq-rgg}--\eqref{eq-pe2}, one concludes that
\begin{equation}\label{eq-c1314}
  c_1=-c_3>0,\quad 0<c_{14}<10^{-5}.
\end{equation}
The constraints on $c_2$ are more complicated, and are given by
\begin{equation}\label{eq-c2b}
  \frac{c_{14}}{1-2c_{14}}<c_2<c_2^u(c_{14}),
\end{equation}
where the upper bound is defined as
\begin{equation}\label{eq-c2u}
  c_2^u(c_{14})=\left\{
  \begin{array}{cc}
    \displaystyle\frac{2(1-4c_{14})}{21}, & 0<c_{14}\lesssim4\times10^{-9}, \\
    &\\
    \displaystyle\frac{c_{14}}{1-2c_{14}-2\times10^{-9}\frac{(2-c_{14})}{c_{14}}}, & 4\times10^{-9}\lesssim c_{14}<10^{-5}.
  \end{array}\right.
\end{equation}
Figure~\ref{fig-c2} shows the constraints on $c_2$ in the range ($0<c_{14}<8\times10^{-9}$), and the shaded region is allowed.
As $c_{14}$ increases, the upper and the lower bounds approach each other.
The bounds on $c_2$ and $c_{14}$ are different from those in Ref.~\cite{Oost:2018tcv} since they used different values for the constraints, such as $|\alpha_1|\le10^{-4}$ and $|\alpha_2|\le10^{-7}$.
Table~\ref{tab-c124s} shows the possible choices for the $c_i$'s such that each column reproduces the corresponding column in Table~\ref{tab-spds}.
These $c_i$'s are of the order of $10^{-9}$, which might still require fine-tuning, albeit less  than when setting $\alpha_1=\alpha_2=0$ and $c_+=1.4\times10^{-15}$.
\begin{table}
  \centering
  \caption{The possible choices for the $c_i$'s to reproduce the speeds in Table~\ref{tab-spds}.
  The last two rows are the speeds of the vector and scalar GWs determined by the choices made in the first three rows.
  The $c_i$'s are normalized by $10^{-9}$.
  }\label{tab-c124s}
  \begin{tabular}{ccccc}
    \hline\hline
    $c_1=-c_3$& & 6.06 & 3.59 & 2.83 \\
    $c_2$ && 3.66 & 2.58 & 2.10 \\
    $c_4$ & & $-4.06$ & $-1.59$ & $-0.83$ \\
    \hline
    $s_v$&& 1.74&1.34&1.19\\
    $s_s$&&1.83&1.29&1.05\\
    \hline
  \end{tabular}
\end{table}
Note that when $c_{13}=0$, the vector polarizations disappear.

\subsection{Pulsar timing arrays}\label{sec-einae-ptas}

A pulsar is a rotating neutron star or a white dwarf with a very strong magnetic field.
It emits a beam of  electromagnetic radiation at a steady rate, and millisecond pulsars can be used as stable clocks \cite{Verbiest:2009kb}.
The presence of GWs will alter the rate, because they will affect the propagation time of the radiation.
This will lead to a change in the time of arrival (TOA), called the timing residual $R(t)$.
Timing residuals are correlated between widely separated pulsars, and the function  $C(\theta)=\langle R_a(t)R_b(t)\rangle$ is used to measure this correlation,
where $\theta$ is the angular separation of pulsars $a$ and $b$, and the brackets $\langle\,\rangle$ indicate the ensemble average over the stochastic GW background.
This underlies the detection of GWs and the probe of the polarization content.
The authors of Refs.~\cite{1975GReGr...6..439E,1978SvA....22...36S,Detweiler:1979wn} considered the effects of GWs in GR on the timing residuals for the first time.
Hellings and Downs proposed a method to detect the effects by cross-correlating the time derivatives of the timing residuals between pulsars \cite{Hellings:1983fr},
while Jenet \textit{et.~al.}  directly used the timing residuals instead of the time derivative \cite{Jenet:2005pv}.
The generalization to massless GWs in alternative metric theories of gravity was soon done in Ref.~\cite{2008ApJ...685.1304L}, and further to massive GWs in Refs.~\cite{Lee:2010cg,Lee:2014awa}.
For work on PTAs, please refer to Refs.~\cite{Chamberlin:2011ev,Yunes:2013dva,Gair:2014rwa,Gair:2015hra} and references therein.

\begin{figure}[h]
\includegraphics[width=0.3\textwidth]{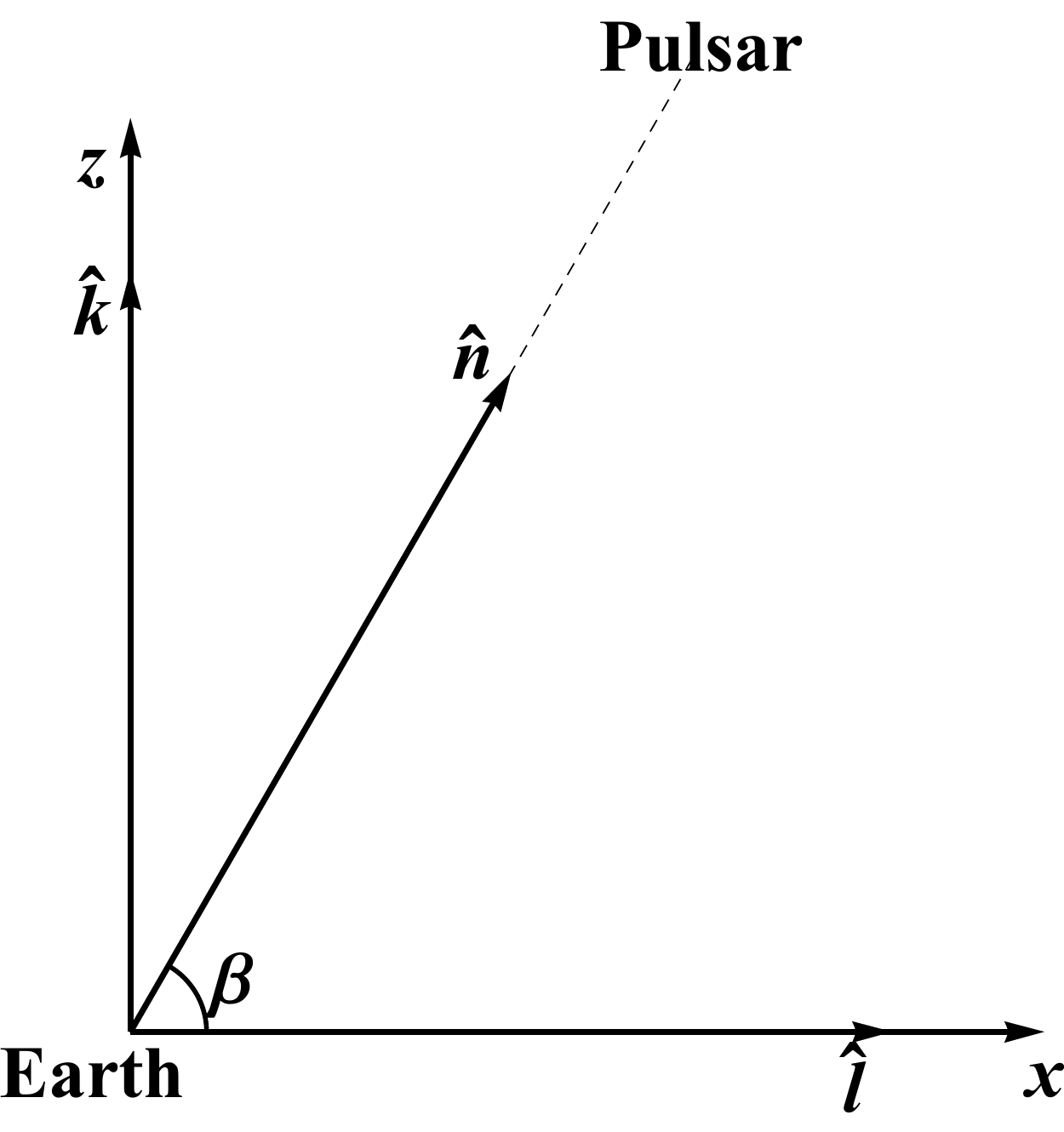}
\caption{The GW propagates in the direction  $\hat k $ and the photon travels in the $-\hat n$ direction at the leading order. $\hat l$ is perpendicular to $\hat k $ and in the same plane determined by $\hat k $ and $\hat n$. The angle between $\hat n$ and $\hat l$ is $\beta$.}\label{fig-coord}
\end{figure}

In order to calculate the timing residual $R(t)$ caused by the GW solution \eqref{gw-plus}--\eqref{gw-scl}, one sets up a coordinate system as shown in Fig.~\ref{fig-coord}.
In this coordinate system, the Earth is at the origin and the distant pulsar is assumed to be stationary at $x_p=(L\cos\beta,0,L\sin\beta)$, when there is no GW.
The GW propagates in the direction $\hat k=(0,0,1)$, and $\hat n$ is the unit vector pointing from the Earth to the pulsar.
Let $\hat l=\hat k \wedge(\hat n\wedge\hat k )/\cos\beta=[\hat n-\hat k (\hat{n}\cdot\hat k )]/\cos\beta$ be the unit vector parallel to the $y$ axis.
At the leading order, i.e., in the absence of GWs,  the photon travels at a 4-velocity $\underline u^\mu=\gamma_0(1,-\cos\beta,0,-\sin\beta)$, where $\gamma_0=\ud t/\ud \lambda$ is a constant and $\lambda$ is an arbitrary affine parameter.
The perturbed photon 4-velocity is $u^\mu=\underline{u}^\mu+V^\mu$.
The photon geodesic equation is
\begin{equation}\label{eq-phogeo}
  \begin{split}
  0=&\frac{\ud u^\mu}{\ud \lambda}+\Gamma^\mu{}_{\rho\sigma}u^\rho u^\sigma\\
  \approx&\gamma_0\frac{\ud V^\mu}{\ud t}+\Gamma^\mu{}_{\rho\sigma}\underline u^\rho \underline u^\sigma.
  \end{split}
\end{equation}
Solving it gives the perturbation in the photon 4-velocity, which is too complicated and will not be reproduced here.

Next, we calculate the 4-velocities of the Earth and the pulsar.
First, we calculate the 4-velocity of the pulsar, which is supposed to be $u_p^\mu=u_p^0(1,\vec v_p)$.
The geodesic equation for the pulsar is,
\begin{equation}\label{accappexp}
\begin{split}
  0=&\frac{\ud^2x^\mu}{\ud\tau^2}+\Gamma^\mu{}_{\rho\nu}\frac{\ud x^\rho}{\ud\tau}\frac{\ud x^\nu}{\ud\tau}\\
   \approx&(u_p^0)^2\left(\frac{\ud^2x^\mu}{\ud t^2}+\Gamma^\mu{}_{00}\right)+u_p^0\frac{\ud u_p^0}{\ud t}\frac{\ud x^\mu}{\ud t},
\end{split}
\end{equation}
where $\tau$ is the proper time.
One sets $x=L\cos\beta$ and $y=0$.
Therefore, the 4-velocity of an observer at rest at the pulsar is
\begin{equation}\label{eq-4v-p}
  u_p^\mu=\left(1+\varphi\cos\omega_s\left(t-\frac{L}{s_s}\cos\beta\right),0,0,-\frac{\varphi}{s_s}\cos\omega_s\left(t-\frac{L}{s_s}\cos\beta\right)\right).
\end{equation}
To get the 4-velocity of an observer at rest at the Earth we simply set $L=0$ in the above expression, so
\begin{equation}\label{eq-4v-e}
  u_e^\mu=\left(1+\varphi\cos\omega_s t,0,0,-\frac{\varphi}{s_s}\cos\omega_s t\right).
\end{equation}
Note that although Einstein-\ae ther theory contains five d.o.f., the velocities of observers (initially at rest) only depend on the scalar d.o.f. $\phi$. In contrast,
the photon's 4-velocity also depends on the tensor and vector d.o.f..

The frequencies measured by the observer at the Earth and by the one at the pulsar are $f_r=-u_\mu u_e^\mu$ and $f_e=-u_\mu u_p^\mu$, respectively.
The relative frequency shift is thus
\begin{equation}\label{eq-fshift}
\begin{split}
  \frac{f_e-f_r}{f_r}=&\frac{(c_{14}-2c_{13})(\hat k\cdot\hat n)^2+s_s^2c_{14}(1-c_{13})}{2(1-c_{13})s_s(s_s+\hat k\cdot\hat n)}[\phi(t,0)-\phi(t-L/s_s,L\hat n)]\\
   &-\frac{c_{13}\hat k\cdot\hat n}{(1-c_{13})(s_v+\hat k\cdot\hat n)}[\hat n\cdot\vec v(t,0)-\hat n\cdot\vec v(t-L/s_v,L\hat n)]\\
   &+\frac{s_g\hat n^j\hat n^k}{2(s_g+\hat k\cdot\hat n)}[h_{jk}^\text{TT}(t,0)-h_{jk}^\text{TT}(t-L/s_g,L\hat n)].
  \end{split}
\end{equation}
This has been put in a coordinate-free form so that this formula always applies regardless of the direction of GW propagation.
The second and last lines both agree with the results in Ref.~\cite{2008ApJ...685.1304L} when $s_g=s_v=1$.
The contribution of the scalar polarization (the first line) does not reduce to the results in Refs.~\cite{Hou:2017bqj,Gong:2017bru}
in a straightforward way where GWs in Horndeski theory are considered, as the scalar fields interact rather differently in these two theories.

In the above discussion, each propagating mode was taken to be monochromatic.
In reality, the stochastic GW background can be described by
\begin{gather}
  \phi(t,\vec x)=\int_{-\infty}^{\infty}\frac{\ud \omega }{2\pi}\int\ud^2\hat k \Big\{\varphi(\omega ,\hat k )\exp[i(\omega t-k \hat k \cdot\vec x)]\Big\},\label{eq-sgw-scl}\\
  \vec v(t,\vec x)=\int_{-\infty}^{\infty}\frac{\ud \omega }{2\pi}\int\ud^2\hat k \Big\{\vec \mu(\omega ,\hat k )\exp[i(\omega t-k \hat k \cdot\vec x)]\Big\},\label{eq-vec-gw}\\
  h_{jk}^\text{TT}(t,\vec x)=\sum_{P=+,\times}\int_{-\infty}^{\infty}\frac{\ud\omega}{2\pi}\int\ud^2\hat k\Big\{\epsilon_{jk}^Ph_P(\omega,\hat k)\exp[i(\omega t-k \hat k \cdot\vec x)]\Big\}, \label{eq-tn-ex}
\end{gather}
where $\varphi(\omega, \hat k), \vec\mu(\omega,\hat k)$, and $h_P(\omega,\hat k)$ are the amplitudes of the scalar, vector, and tensor GWs oscillating at $\omega$ and propagating in the direction  $\hat k$, respectively.
$\epsilon^P_{jk}$ is the polarization matrix and $P=+,\times$.
$\vec \mu(\omega,\hat k)$ is transverse, i.e., $\hat k\cdot\vec\mu=0$.
So if the unit vectors $\hat e_{\tilde 1}$, $\hat e_{\tilde 2}$, and $\hat e_{\tilde 3}=\hat k$ form a triad such that $\hat e_{\tilde j}\cdot\hat e_{\tilde l}=\delta_{\tilde j\tilde l}$, and $\hat e_{\tilde 3}=\hat e_{\tilde 1}\times\hat e_{\tilde 2}$,
then $\vec\mu(\omega,\hat k)$ has two d.o.f. which can be expressed as
\begin{equation}\label{eq-sp-vec-d.o.f.}
  \vec\mu(\omega,\hat k)=\hat e_{\tilde 1}\mu_{\tilde 1}(\omega,\hat k)+\hat e_{\tilde 2}\mu_{\tilde 2}(\omega,\hat k).
\end{equation}

\begin{figure}
  \centering
  \includegraphics[width=0.5\textwidth]{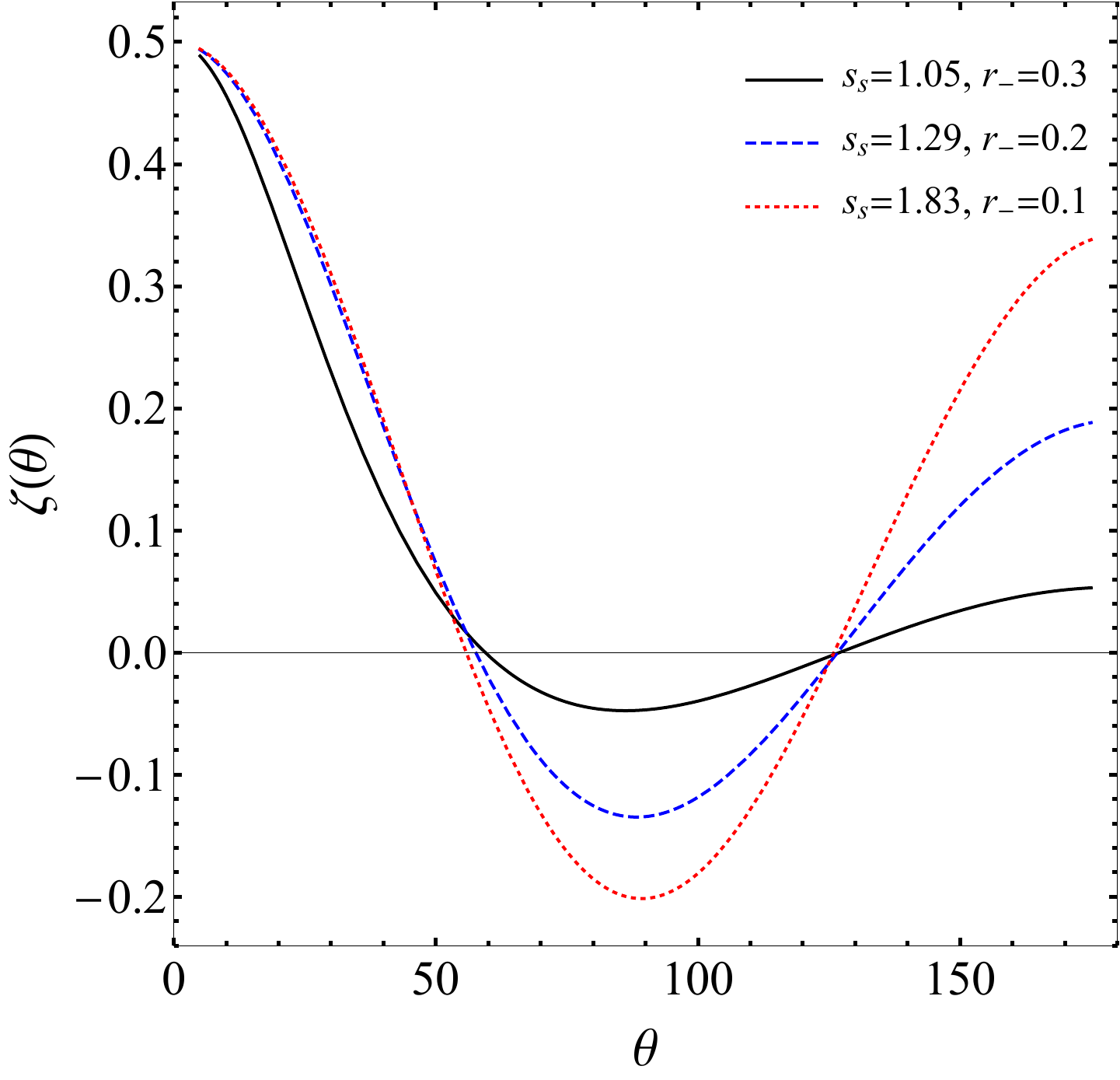}
  \caption{The normalized cross correlation $\zeta(\theta)=C_s(\theta)/C_s(0)$ for the scalar GW.
  $\zeta(\theta)$ is plotted for different propagation speeds corresponding to different $r_{-}$ [see Eq.~\eqref{eq-cminus}].}\label{fig-zeta-scl}
\end{figure}

\begin{figure}
  \centering
  \includegraphics[width=0.5\textwidth]{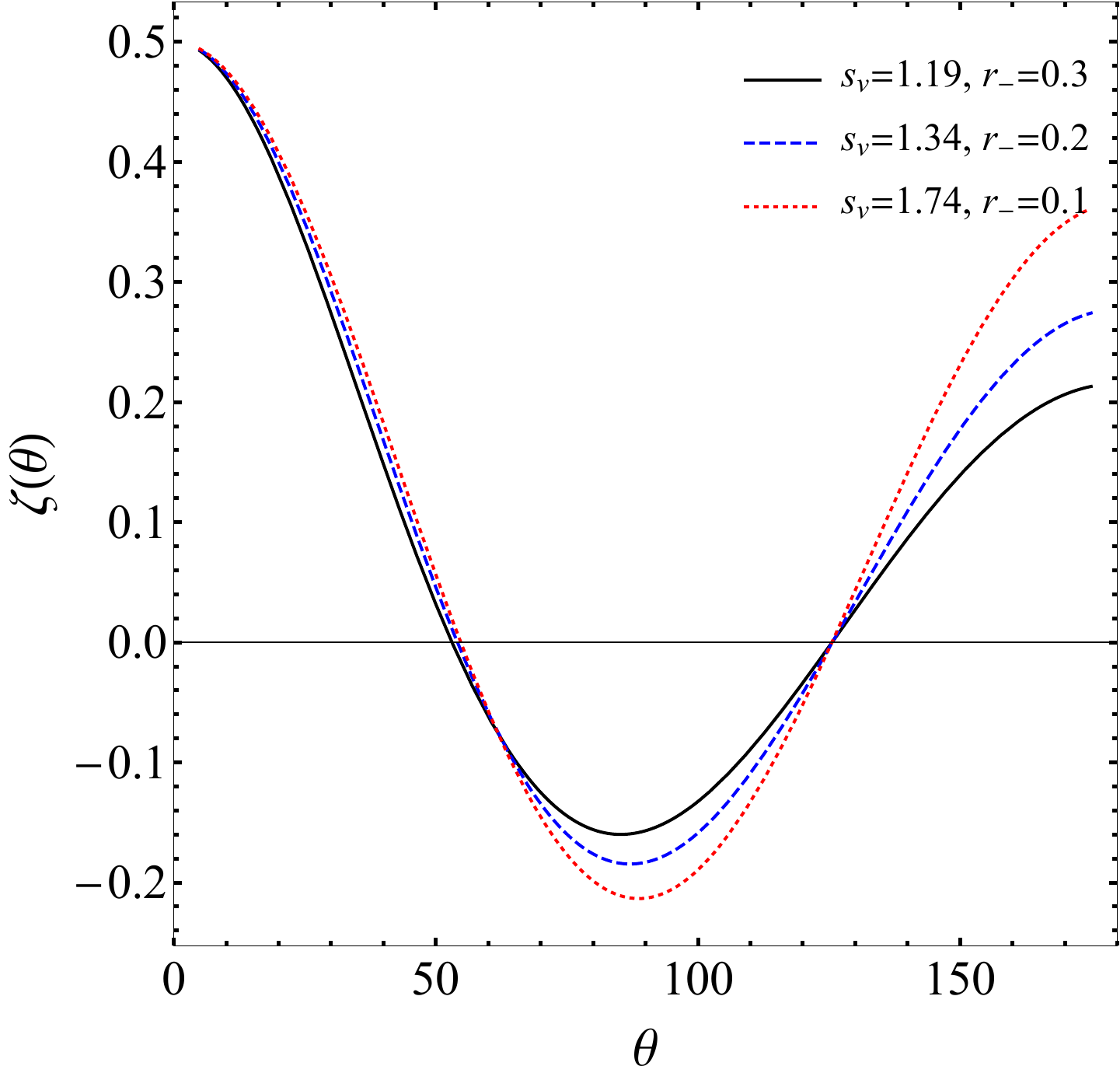}
  \caption{The normalized cross correlation $\zeta(\theta)=C_v(\theta)/C_v(0)$ for the vector GW. $\zeta(\theta)$ is plotted for different propagation speeds corresponding to different $f$'s (see Eq.~\eqref{eq-cminus}).}\label{fig-vec-ncors}
\end{figure}

\begin{figure}
  \centering
  \includegraphics[width=0.5\textwidth]{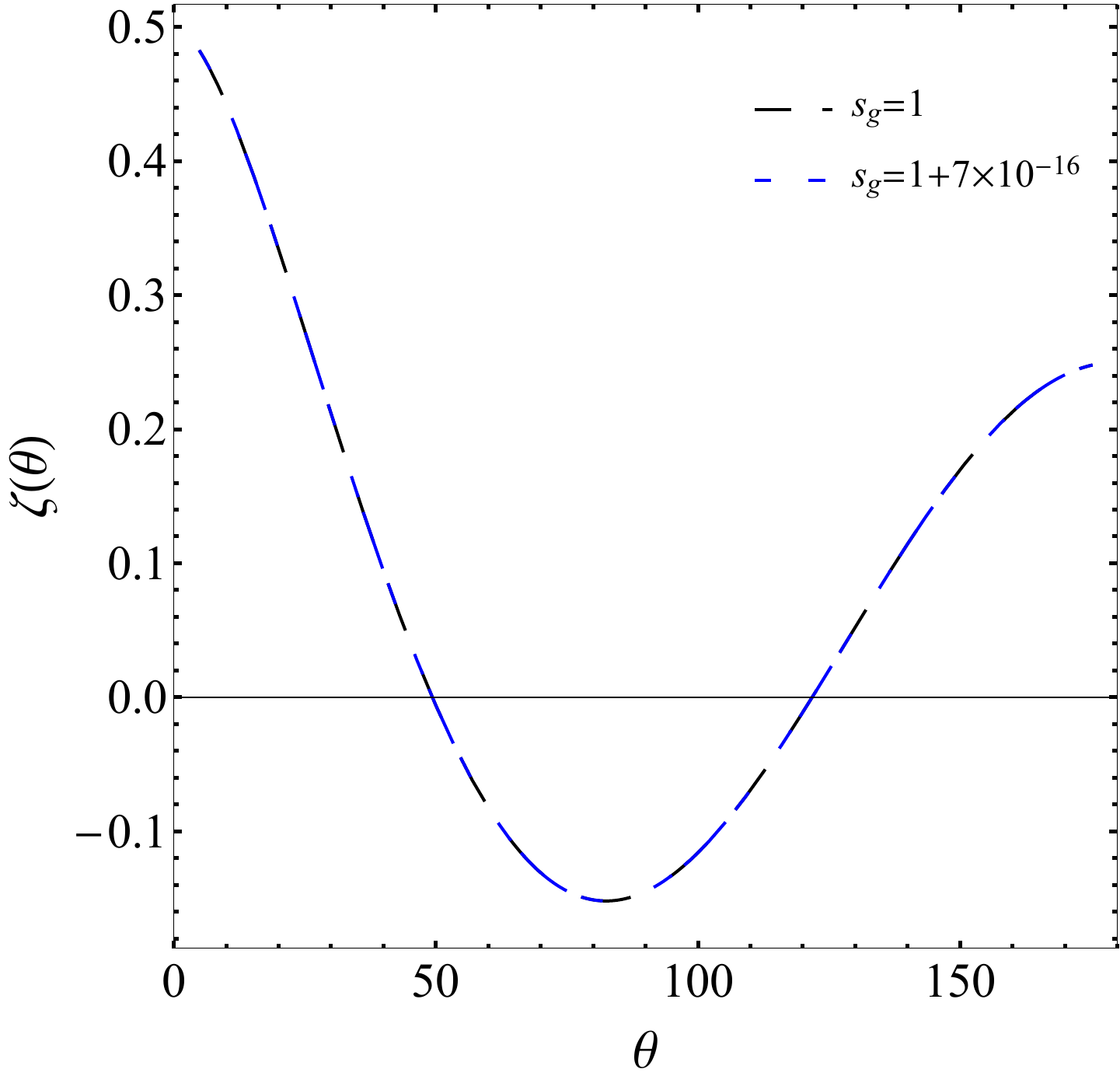}
  \caption{The normalized cross correlation $\zeta(\theta)=C_g(\theta)/C_g(0)$ for the tensor GW at $s_g=1+7\times10^{-16}$. Also shown is $\zeta(\theta)$ for GR ($s_g=1$). Since the difference in the speeds is extremely small, the two curves nearly overlap with each other.}\label{fig-zeta-tns}
\end{figure}

Integrating the relative frequency shift gives the timing residual
\begin{equation}\label{rtsl}
  R(T)=\int_{-\infty}^{\infty}\frac{\ud \omega }{2\pi}\int\ud^2\hat k \int_{0}^{T}\ud t\frac{f_e-f_r}{f_r},
\end{equation}
where the argument $T$ is the total observation time.
Suppose that the stochastic GW background is isotropic, stationary, and independently polarized; then, one defines the characteristic strains $\varphi_c(\omega), \mu^c_{\tilde j}(\omega)$, and $h_c^P(\omega )$ in the following manner:
\begin{gather}
\langle \varphi^*(\omega ,\hat k )\varphi(\omega' ,\hat k' )\rangle=\delta(\omega -\omega' )\delta^{(2)}(\hat k -\hat k' )\frac{|\varphi_c(\omega )|^2}{\omega },\label{defvc}\\
  \langle\mu^*_{\tilde j}(\omega,\hat k)\mu_{\tilde l}(\omega',\hat k')\rangle=\delta(\omega-\omega')\delta^{(2)}(\hat k-\hat k')\delta_{\tilde j\tilde l}\frac{|\mu^c_{\tilde j}(\omega)|^2}{\omega},\label{eq-def-ema-vec}\\
  \langle h^{*}_P(\omega ,\hat k)h_P(\omega ',\hat k')\rangle=\delta(\omega -\omega ')\delta^{(2)}(\hat k-\hat k')\delta^{PP'}\frac{\pi|h_c^P(\omega )|^2}{4\omega }  \label{eq-def-s-h}
\end{gather}
where a star $*$ indicates  complex conjugation.
The characteristic strains are proportional to $\omega^\alpha$, where $\alpha$ is the power-law index.
The cross-correlation function $C(\theta)=\langle R_a(T)R_b(T)\rangle$ can thus be obtained.
The detailed calculation has been relegated to the Appendix~\ref{sec-app-cc}.
The normalized cross correlation $\zeta(\theta)=C(\theta)/C(0)$ is calculated numerically, and
the results are shown in Figs.~\ref{fig-zeta-scl}, \ref{fig-vec-ncors}, and \ref{fig-zeta-tns} for the scalar, vector, and tensor polarizations, respectively.

Figure~\ref{fig-zeta-scl} shows  the behavior of $\zeta(\theta)$ as a function of $\theta$ at different speeds $s_s$,
corresponding to different $r_{-}$ [see Eq.~\eqref{eq-cminus}] for the scalar polarization.
As one can see, $\zeta(\theta)$ increases with $\theta$ in the small- and large-angle ranges, while it decreases in the intermediate-angle range.
It becomes negative in certain ranges.
The inspection of the dependence of $\zeta(\theta)$ on $s_s$ or $r_{-}$ shows that $\zeta(\theta)$ is more sensitive to $s_s$ or $r_{-}$ when $\theta$ is large.
As discussed in the Appendix~\ref{sec-app-cc}, $\zeta(\theta)$ does not depend on the power-law index $\alpha$.
The behavior of $\zeta(\theta)$ in this work differs greatly from that for the scalar GWs in the scalar-tensor theory obtained in Refs.~\cite{Gong:2017bru,Hou:2017bqj},
where $\zeta(\theta)$ for the scalar GWs in Horndeski theory was obtained for different masses and the power-law index $\alpha$,
and it is always positive and a decreasing function of $\theta$ \cite{Hou:2017bqj}.
The behavior of $\zeta(\theta)$ in this work is also different from that for the transverse breathing and  longitudinal polarizations  presented in Refs.~\cite{2008ApJ...685.1304L,Lee:2010cg,Lee:2014awa}, where these two polarizations were treated as independent of each other.

Figure~\ref{fig-vec-ncors} shows how $\zeta(\theta)$ varies as a function of $\theta$ at different $s_v$ or $r_{-}$ for the vector polarizations.
One finds that $\zeta(\theta)$ also has similar behavior as that for the scalar GWs and it does not depend on the power-law index $\alpha$,
but it is not as sensitive to $s_v$ or $r_{-}$ as the one for the scalar GWs.
Comparing this figure with the bottom-left panel in Fig.~1 in Ref.~\cite{2008ApJ...685.1304L} shows that $\zeta(\theta)$ becomes flatter at large angles in Ref.~\cite{2008ApJ...685.1304L}.
$\zeta(\theta)$ for the massive GWs was considered in Ref.~\cite{Lee:2014awa}, and the bottom-left panel in Fig. 1 in Ref.~\cite{Lee:2014awa} is for the vector polarizations.
They show some similarities to the one in the current work.

Figure~\ref{fig-zeta-tns} shows $\zeta(\theta)$ for the tensor polarizations at $s_g=1+7\times10^{-16}$.
Also shown is the one for GR labeled by $s_g=1$, which is given by \cite{Hellings:1983fr,2008ApJ...685.1304L}
\begin{equation}\label{eq-ncc-gr}
  \zeta(\theta)=\frac{3}{4}(1-\cos\theta)\ln\frac{1-\cos\theta}{2}+\frac{1}{2}-\frac{1-\cos\theta}{8}+\frac{\delta(\theta)}{2}.
\end{equation}
Since the difference in the speeds is extremely small, the two curves nearly overlap with each other.

\begin{figure}
  \centering
  \includegraphics[width=0.5\textwidth]{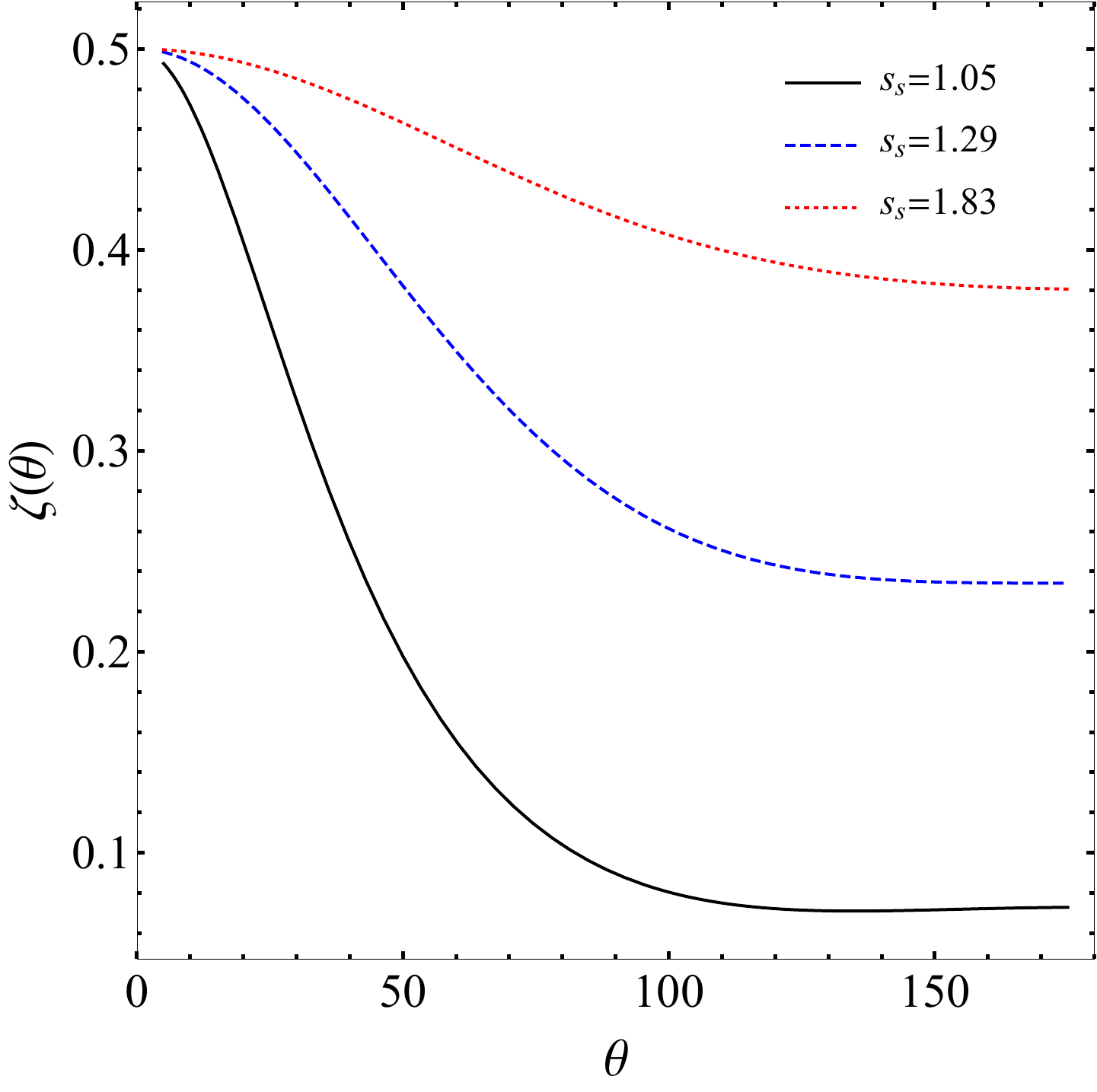}
  \caption{The normalized cross-correlation function $\zeta(\theta)=C_s(\theta)/C_s(0)$ for the scalar GW when the $c_i$'s take values in Table~\ref{tab-c124s}.}\label{fig-scl-c130}
\end{figure}

{If one chooses the values for the $c_i$'s given in Table~\ref{tab-c124s}, the normalized cross-correlation function $\zeta(\theta)$ for the scalar GW is modified, as shown in Fig.~\ref{fig-scl-c130}.
$\zeta(\theta)$ for the tensor GW is described by the curve labeled by ``$s_g=1$" in Fig.~\ref{fig-zeta-tns}.
Since when $c_{13}=0$, the vector polarizations disappear, we do not plot the corresponding cross-correlation functions.
It is clear that $\zeta(\theta)$ for the scalar GW behaves rather differently than the one for the tensor GW.}

Finally, let us compare the cross-correlation functions for the scalar, vector, and tensor polarizations in Einstein-\ae ther theory.
If one chooses the $c_i$'s to make $\alpha_1=\alpha_2=0$, the cross-correlation functions for the vector modes are quite similar to those for the tensor modes with a small variation depending on the speed $s_v$, as shown in Figs.~\ref{fig-zeta-scl}, \ref{fig-vec-ncors}, and \ref{fig-zeta-tns}.
The cross-correlation function for the scalar mode is somewhat different than those for the vector and the tensor modes when its speed is small, for example, $s_s=1.05$ (the black curve in Fig.~\ref{fig-zeta-scl}), but when its speed is larger the difference becomes smaller.
Compared to the results in Refs.~\cite{2008ApJ...685.1304L,Lee:2010cg,Lee:2014awa,Hou:2017bqj},
Figs.~\ref{fig-zeta-scl}, \ref{fig-vec-ncors}, and \ref{fig-zeta-tns} show greater similarities among the cross-correlation functions for different polarizations, so it is more difficult to use PTAs to distinguish different polarizations and thus test whether extra polarizations exist  in Einstein-\ae ther theory.
{However, if one chooses  the $c_i$'s to make $s_g=1$ (i.e., the  values in Table~\ref{tab-c124s}), there is only one extra polarization state, and its cross-correlation function differs from that of the tensor modes greatly.
So it would be easier to use PTAs to distinguish different polarizations in Einstein-\ae ther theory, and thus falsify it if no extra polarization is observed.
}

\section{Gravitational Waves in the generalized TeVeS Theory}\label{sec-gtvs}

The action of the generalized TeVeS theory is given by the sum of that of Einstein-\ae ther theory \eqref{aeact}
and the one for the additional scalar field $\sigma$,
\begin{equation}\label{eq-act-sig}
  S_\sigma=-\frac{8\pi}{\jmath^2\ell^2G}\int\ud^4x\sqrt{-g}\mathcal F(\jmath\ell^2 j^{\mu\nu}\sigma_{,\mu}\sigma_{,\nu}),
\end{equation}
where $j^{\mu\nu}=g^{\mu\nu}-u^\mu u^\nu$,  $\jmath$ is a dimensionless positive parameter, and $\ell$ is a constant with dimensions of length.
The function $\mathcal F$ is dimensionless and chosen to produce the relativistic MOND phenomena.
Here, since the action of the vector field is that of the \ae ther, we simply use $u^\mu$ to represent $\mathfrak U^\mu$.

Because of the extra scalar field $\sigma$, the equations of motion \eqref{eq-ae-1} and \eqref{eq-ae-2} are modified.
First, on the right-hand side of Eq.~\eqref{eq-ae-1} one has to add the contribution $\tau_{\mu\nu}/2$ of the stress-energy tensor of the scalar field $\sigma$, which is
\begin{equation}\label{eq-st-sig}
  \tau_{\mu\nu }=\frac{16\pi\mathcal F'(y)}{\jmath}(\sigma _{,\mu}\sigma _{,\nu }-2u^\mu\sigma _{,\mu}u_{(\mu}\sigma _{,\nu )})-g_{\mu\nu }\frac{8\pi\mathcal F(y)}{\jmath ^2\ell^2},
\end{equation}
where $y=\jmath\ell^2 j^{\mu\nu}\sigma_{,\mu}\sigma_{,\nu}$ and $\mathcal F'(y)=d\mathcal F(y)/dy$.
Second, one has to add $-\frac{8\pi}{\jmath}\mathcal F'(y)u^{\nu}\sigma_{,\nu}g^{\mu\rho }\sigma_{,\rho }$ to the right-hand side of Eq.~\eqref{eq-ae-2}.
Finally, the equations of motion for the scalar field $\sigma$ are
\begin{equation}\label{eq-sig-eom}
\nabla_\nu[\mathcal F'(y)j^{\mu\nu}\sigma_{,\mu}]=0.
\end{equation}

Another important difference between Einstein-\ae ther theory and the generalized TeVeS theory is that there are two metric tensors in the latter.
The first metric $g_{\mu\nu}$ appearing in the actions \eqref{aeact} and \eqref{eq-act-sig} is called the ``Einstein metric."
The second metric $\tilde g_{\mu\nu}=e^{-2\sigma}g_{\mu\nu}-2u_\mu u_\nu\sinh(2\sigma)$ is the physical metric, and the matter fields $\psi_m$ minimally couple to this metric, i.e., the matter action is symbolically given by
\begin{equation}\label{eq-mact}
  S'_m=\int\ud^4\sqrt{-\tilde g}\mathcal L(\tilde g_{\mu\nu},\psi_m,\tilde\nabla_\mu\psi_m),
\end{equation}
where $\tilde\nabla_\mu$ is the covariant derivative compatible with $\tilde g_{\mu\nu}$.
Therefore, a neutral test particle travels on the geodesic determined by $\tilde g_{\mu\nu}$ in free fall.
In general, the geodesics of $g_{\mu\nu}$ differ from those defined by $\tilde g_{\mu\nu}$, unless $\sigma=0$.

\subsection{Gravitational-wave solutions}\label{sec-gtvs-gw}

In this work we find the GW solutions in the flat spacetime background.
The background solution
\begin{equation}\label{eq-sol-fla}
  g_{\mu\nu}=\eta_{\mu\nu},\quad u^\mu=\underline u^\mu,\quad \sigma=\sigma_0
\end{equation}
(where $\sigma_0$ is a constant) requires that $\mathcal F(0)=0$.
Now, we perturb $g_{\mu\nu}$ and $u^\mu$ according to Eq.~\eqref{eq-pbs}, and the scalar field $\sigma$ is perturbed in the following way:
\begin{equation}\label{eq-perbs}
  \sigma=\sigma_0+\varsigma.
\end{equation}
The linearized Einstein equation and the vector equation take the exact same forms as in Einstein-\ae ther theory, which have been solved in Sec.~\ref{sec-eom-ginv}.
The linearized scalar equation is
\begin{equation}\label{eq-scl-li}
  \partial_\nu[\mathcal F'(0)j^{\mu\nu}_0\varsigma_{,\mu}]=0,
\end{equation}
with $j^{\mu\nu}_0=\eta^{\mu\nu}-\underline{u}^\mu\underline u^\nu=\text{diag}(-2,1,1,1)$.
If one chooses the original form for $\mathcal F$ \cite{Bekenstein:2004ne}, $\mathcal F''(0)$ blows up.
However, there are other choices for $\mathcal F$ as given in Ref.~\cite{Skordis:2009bf}, such that $\mathcal F''(0)$ is finite \footnote{For example, one can set $\mu_a=1$ and $n=3$ in Eq.~(38) in Refs.~\cite{Bourliot:2006ig,Skordis:2009bf}, so that $\mathcal F'(y)\approx \mu_0\left(3-\frac{64\pi \ell^2}{27\mu_0^2}y\right)+O(y^2)$.}.
Expanding the above relation \eqref{eq-scl-li} gives
\begin{equation}\label{eq-scl-eom-li}
  -\ddot\varsigma+\frac{1}{2}\nabla^2\varsigma=0,
\end{equation}
so the scalar perturbation $\varsigma$ propagates at the speed  $s_0=1/\sqrt{2}$.
Therefore, a plane-wave solution propagating in the positive $z$ direction is
\begin{equation}\label{eq-scl-sol}
  \varsigma=\varsigma_0\cos[\omega(t-z/s_0)],
\end{equation}
where $\varsigma_0$ is the amplitude and $\omega$ is the angular frequency.
The plane-wave solutions for the metric and the vector fields have been given in Eqs. \eqref{gw-plus}--\eqref{gw-scl}.

Up to the linear order, the physical metric is thus
\begin{gather}
\tilde{g}_{00}=e^{2\sigma_0}(-1+h_{00}-2\varsigma), \\
\tilde{g}_{0j}=2v^j \sinh(2\sigma_0), \\
\tilde{g}_{jk}=e^{-2\sigma_0}[\delta_{jk}(1-2\varsigma)+h_{jk}].
\end{gather}
Note that this metric is written in  coordinates determined by the Einstein metric $g_{\mu\nu}$,
and the gauge conditions $h_{0j}=0$ and $\partial_jv^j=0$ have been imposed.
If one performs the  coordinate transformation \cite{Sagi:2010ei}
\begin{equation}\label{eq-cort}
  \tilde{x}^{0}=e^{\sigma_0}x^0,\quad \tilde{x}^{j}=e^{-\sigma_0}x^j,
\end{equation}
the physical metric becomes
\begin{gather}
\tilde{g}_{00}=-1+h_{00}-2\varsigma , \\
\tilde{g}_{0j}=2v^j \sinh(2\sigma_0), \\
\tilde{g}_{jk}=\delta_{jk}(1-2\varsigma )+h_{jk}.
\end{gather}
Note that all of the fields on the right-hand side in the above expressions are written as functions of $\tilde{x}^{0}$ and $\tilde{x}^{j}$ implicitly.
In this coordinate system, the speeds become
\begin{gather}
\tilde s_g^2=\frac{e^{-4\sigma_0}}{1-c_{13}},\label{eq-teves-sg2}\\
\tilde s_v^2=e^{-4\sigma_0}\frac{c_1-c_1^2/2+c_3^2/2}{c_{14}(1-c_{13})},\label{eq-teves-sv2}\\
\tilde s_s^2=\frac{e^{-4\sigma_0}c_{123}(2-c_{14})}{c_{14}(1-c_{13})(2+2c_2+c_{123})},\label{eq-teves-ss2}\\
\tilde s_0^2=\frac{e^{-4\sigma_0}}{2}.\label{eq-teves-s02}
\end{gather}
Again, the speeds are not necessarily 1, and are generally different from one another.
When all speeds are 1, the following conditions should be satisfied:
\begin{equation}\label{eq-s1-c}
  \sigma_0=-\frac{\ln 2}{4},\quad c_1=c_4-\frac{1}{2},\quad c_3=-c_4-\frac{1}{2},\quad c_2=\frac{1}{2(1-2c_4)}.
\end{equation}
However, a negative $\sigma_0$ is not acceptable in this theory \cite{Sagi:2009kd}.

In total there are six d.o.f.: in addition to those that resemble the five d.o.f. in Einstein-\ae ther theory, there is one more scalar d.o.f., $\sigma$.
Note that there are two scalar d.o.f., $\sigma$ and $\Omega$, in this theory.
All of these d.o.f. will affect the polarization content of GWs in the generalized TeVeS theory.
The polarization content is obtained by calculating the linearized geodesic deviation equation
$\ddot{\tilde{x}}^{j}=-\tilde R_{\tilde t\tilde j\tilde t\tilde k}\tilde{x}^{k}$, where $\tilde R_{\tilde t\tilde j\tilde t\tilde k}$
is the linearized Riemann tensor calculated using the physical metric $\tilde g_{\mu\nu}$.
There are six polarization states in the generalized TeVeS theory:
the plus polarization $\hat P_+=-\tilde R_{\tilde t\tilde x\tilde t\tilde x}+\tilde R_{\tilde t\tilde y\tilde t\tilde y}=\ddot h_+$
and the cross polarization $\hat P_\times=\tilde R_{\tilde t\tilde x\tilde t\tilde y}=-\ddot h_\times$;
the vector-$x$ polarization $\hat P_{xz}=\tilde R_{\tilde t\tilde x\tilde t\tilde z}
=-\{c_{13}(1+2\sinh[2\sigma_0)]-2\sinh(2\sigma_0)\}\ddot v_1/[2(1-c_{13})\tilde s_v]$,
and the vector-$y$ polarization $\hat P_{yz}=\tilde R_{\tilde t\tilde x\tilde t\tilde y}
=-\{c_{13}[1+2\sinh(2\sigma_0)]-2\sinh(2\sigma_0)\}\ddot v_2/[2(1-c_{13})\tilde s_v]$;
the transverse breathing polarization $\hat P_b=\tilde R_{\tilde t\tilde x\tilde t\tilde x}+\tilde R_{\tilde t\tilde y\tilde t\tilde y}=-c_{14}\ddot\phi+2\ddot\varsigma$,
and the longitudinal polarization
$$\hat P_l=\tilde R_{\tilde t\tilde z\tilde t\tilde z}=\left[\frac{(c_{14}-2c_{13})}{2(c_{13}-1)\tilde s_s^2}-\frac{c_{14}}{2}\right]\ddot\phi+\left(1+\frac{1}{\tilde s_0^2}\right)\ddot\varsigma.$$
Therefore, the scalar d.o.f., $\phi$ and $\varsigma$, excite two mixed states of $\hat P_b$ and $\hat P_l$.
As in Einstein-\ae ther theory, none of the Newman-Penrose variables  vanish in general.

\subsection{Discussion on the constraints}\label{sec-teves-cons}

Sagi calculated the post-Newtonian parameters for the generalized TeVeS theory \cite{Sagi:2009kd},
and $\alpha_1$ and $\alpha_2$ are given in Eqs.~(46)--(48) in Ref. \cite{Sagi:2009kd}, which are too complicated to be reproduced here.
In her equations, $K=(c_1-c_3)/2,\,K_+=c_{13}/2,\,K_2=c_2,$ and $K_4=-c_4$.
She also found that
\begin{equation}\label{eq-teves-gcs}
  G=G_\text{N}\frac{4\pi(2-c_{14})}{8\pi+\jmath(2-c_{14})},
\end{equation}
which should be positive (where $G_\text{N}$ is Newton's constant).
{Using the expressions for $\alpha_1$ and $\alpha_2$, one can solve for $\jmath$ and $c_2$ in terms of $\sigma_0$, $c_j$ ($j\ne2$), and the $\alpha$'s.
Note that $\alpha_1$ and $\alpha_2$ are not necessarily set to zero in the following discussion.
}

Next, the observations of GW170817 and GRB 170817A set bounds on the propagation speed of the tensor mode.
The above discussion shows that there are four different speeds for different polarizations.
Here, we set $\tilde s_g=1+\delta$ with $-3\times10^{-15}<\delta<7\times10^{-16}$.
This is the third constraint for this theory, and it relates $\sigma_0$ to $c_{13}$.
Therefore, the parameter space reduces to three dimensions, conveniently parametrized by $c_1, c_3$, and $c_4$.

In addition, the MOND effects should not be too large in the Solar System, which requires that $\jmath$ is of the order of 0.01 \cite{Bekenstein:2004ne,Sagi:2009kd}.
Finally, by studying the neutron star and black hole solutions, the authors of Refs.~\cite{Sagi:2007hb,Lasky:2008fs,Lasky:2009sw} set a new bound, i.e., $c_{14}\lesssim 1$.
With these constraints and bounds, one can scan the reduced parameter space to search for the parameter ranges such that all speeds are of the order of unity.
The strategy is given below:
\begin{enumerate}
  \item Start with a relatively larger reduced parameter space $S_0$, i.e., $-10<c_1, c_3,c_4<10$, and search for the subspace $S_1$ such that $\tilde s_v$ and $\tilde s_s$ are smaller than an upper bound $v_0$ (say, $10^{13}$) with a common step size $\Delta^{(0)}=20/N$, where $N$ is an integer.
      In this search, all of the constraints and bounds should be taken into account.
  \item If such a subspace $S_1$ is found, one proceeds to the next iteration. In this iteration, the reduced parameter space is $S_1$ and the step size for $c_i$ is given by $\Delta_i^{(1)}=\delta c_i/N\,(i=1,3,4)$, where $\delta c_i$ is the difference between the maximum and minimum values of $c_i$ that define $S_1$.
      The new speed bound $v_1$ is also updated, given by the minimum speed $\tilde s_v$ or $\tilde s_s$ found in the previous iteration.
  \item If such a subspace $S_1$ cannot be found, the iteration terminates.
\end{enumerate}
One repeats the above steps until one cannot find a subspace $S_n$ such that $\tilde s_v,\,\tilde s_s<v_n$ in this subspace after $n$ iterations.
In order to avoid the influence of the step sizes on the final result, one can vary $N$.
It turns out that one cannot find such a subspace in which $\tilde s_v$ and $\tilde s_s$ are both of the order of unity, while all of the constraints and bounds are satisfied simultaneously.
This  can be understood roughly by expressing $\tilde s_v,\,\tilde s_s$ in terms of $\tilde s_g$ with $\alpha_1=\alpha_2=0$,
\begin{gather}
\tilde s_v^2=\frac{\tilde s_g^2}{2}\frac{2c_1-c_1^2+c_3^2}{c_{14}}\\\
\approx\frac{1}{2}\frac{2c_1-c_1^2+c_3^2}{c_{14}},\label{eq-sv-a120}\\
\tilde  s_s^2= \frac{4\tilde s_g^2}{3}\left[1-\frac{c_{14}}{2(1-\tilde{s}_g^{-2})}\right]^2\frac{\tilde s_v^2}{2-c_{14}}\nonumber\\
  \approx  \frac{4}{3}\left(1-4\delta^{-1}c_{14}\right)^2\frac{\tilde s_v^2}{2-c_{14}}.\label{eq-ss-a120}
\end{gather}
At the same time, $\jmath$ can be approximated as
\begin{equation}\label{eq-j-app}
  \jmath\approx\frac{4\pi c_{14}}{c_{14}-2}+8\pi\delta,
\end{equation}
so $c_{14}$ is of the order of $10^{-2}$.  If  $\tilde s_v$ is of the order of unity and $\delta$ takes the largest value $|\delta|\sim 10^{-15}$, $\tilde s_s$ is of order $10^{13}$!
Any attempt to reduce $\tilde s_s$ to be of the order of unity while keeping $\tilde s_v\sim 1$ fails.
A more serious problem is that, $\tilde s_s$ blows up as $\delta$ approaches 0 as one can check from Eq.~\eqref{eq-ss-a120}.
{On the other hand, one may also consider  simply setting $\delta=0$ (i.e., $\tilde s_g=1$) without requiring $\alpha_1=\alpha_2=0$.
In this case, one obtains that
\begin{equation}\label{eq-j-app-2}
  \jmath =\frac{8\pi(\alpha_1+4c_{14})}{(8+\alpha_1)(c_{14}-2)},
\end{equation}
which can be solved for $c_{14}$.
At the same time, one finds  that
\begin{equation}\label{eq-ss2-app-2}
\begin{split}
  \tilde s_s^2&=\frac{(8+\alpha_1)c_{14}}{7\alpha_1(2-c_{14})}\\
  &=-\frac{(\alpha_1+8)\jmath+4\pi\alpha_1}{28\pi\alpha_1}\\
  &\sim 10^2.
\end{split}
\end{equation}
So the scalar field $\phi$ will still propagate at a large (although not necessarily infinite) speed, which might lead to a faster decay of the orbit  of a binary system.
}

\begin{figure}
  \centering
  \includegraphics[width=0.5\textwidth]{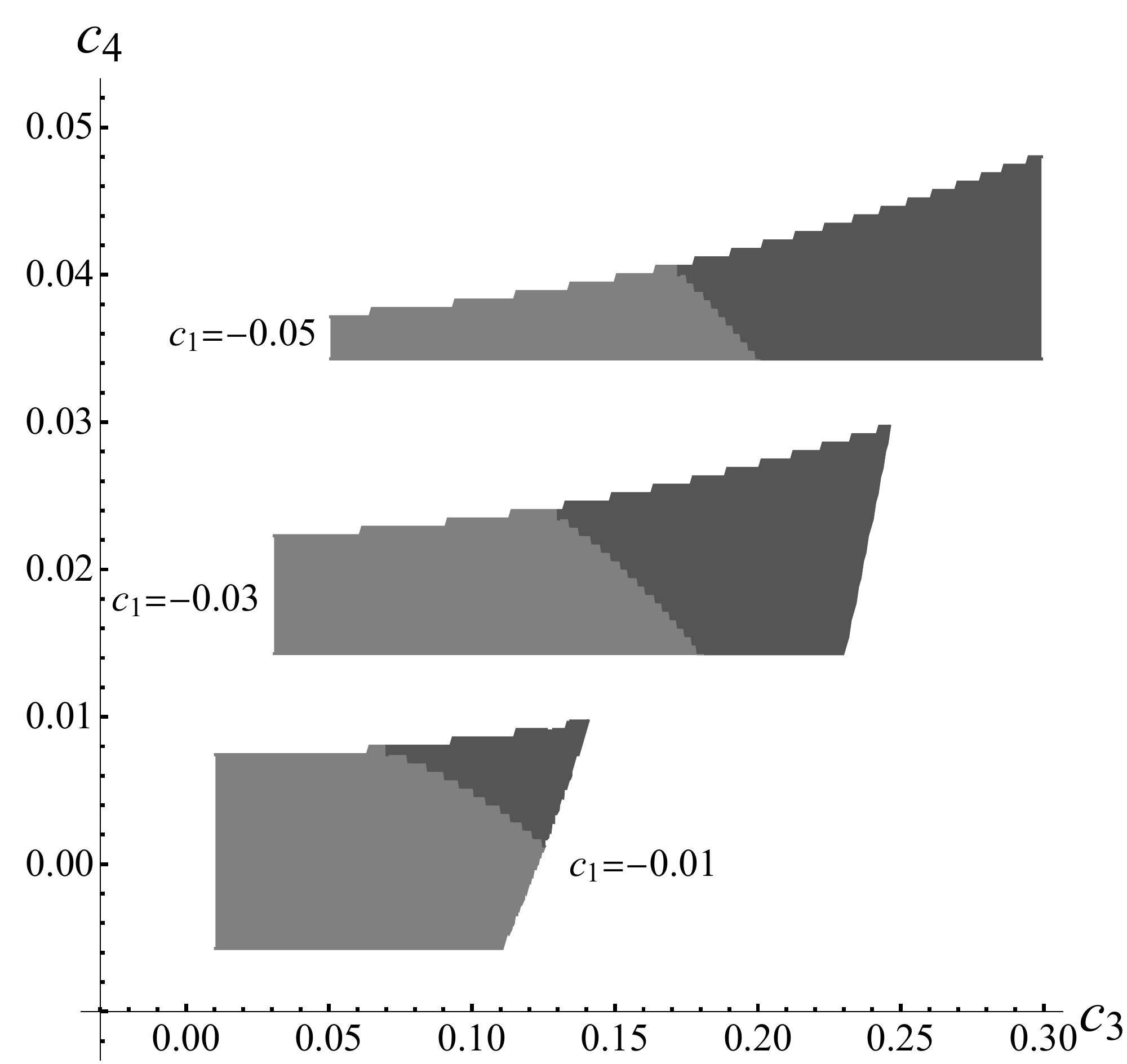}
  \caption{Parameter subspaces (colored areas) allowed by the experimental constraints.
  There are three chunks of allowed parameter subspaces corresponding to different values of $c_1$.
  Each allowed region is divided into two pieces.
  The gray areas represent the parameter subspaces in which the strong coupling problem does not exist,
  while the dark gray areas represent the parameter subspaces in which the strong coupling problem does exist.
  (The jagged boundaries are due to the finite step size used in the scanning.)}\label{fig-sc-nosc}
\end{figure}

A very large speed might lead to the strong coupling problem, and the scalar mode $\phi$ might not be excited.
In this case, one has to integrate out this mode and then apply the experimental constraints to the resulting theory.
In order to examine whether the strong coupling problem arises, one needs to expand the action up to the cubic order in the scalar perturbations, and calculate all of the coefficients of the terms in the cubic action after canonically normalizing the scalar d.o.f..
The resulting cubic Lagrangian is very complicated and will not be presented here.
It shows that the strong coupling problem exists in some parameter subspaces.
For example, Fig.~\ref{fig-sc-nosc} shows the allowed parameter subspaces, which were obtained by scanning the parameter space.
The gray areas represent the parameter subspaces in which the strong coupling problem does not exist,
while the dark gray areas represent the parameter subspaces in which the strong coupling problem does exist.
So in these dark gray areas the above analysis on the scalar mode cannot be applied.
These allowed parameter subspaces depend on $\delta$ and the $\alpha_i$'s.
However, the changes due to varying $\delta$ and the $\alpha_i$'s are very small.
So the generalized TeVeS theory is excluded due to the large or even infinite speed $\tilde s_s$, given the speed limits on the tensor GW mode, in the parameter space where the strong coupling problem does not exist.

\section{Conclusion}\label{sec-con}

In this work,  we discussed the linear GW solutions around the flat spacetime background and the
polarization contents of Einstein-\ae ther theory and the generalized TeVeS theory.
It turns out that both theories predict the existence of tensor, vector, and scalar GWs, each propagating with different speeds, generally different from 1.
In obtaining the GW solutions, we used the gauge-invariant variables to help separate the physical d.o.f..
There are five polarization states in Einstein-\ae ther theory,
while the generalized TeVeS theory predicts the existence of six polarization states.
The transverse breathing mode is mixed with the longitudinal mode to form a single state for the scalar polarization in Einstein-\ae ther theory.
The two scalar polarizations in the generalized TeVeS theory are two mixed states of the transverse breathing and longitudinal modes.
In addition, the possible experimental tests of the polarizations in Einstein-\ae ther theory
have been considered by using the cross-correlation functions of PTAs for the various polarizations together with the speed bounds on GWs set by the observations of GW170817 and GRB 170817A.
We found that the cross-correlation functions for different polarizations look very similar to each other in some parameter regions,
and this means that it will be difficult for PTAs to identify the polarizations.
{ However, in the parameter regions with $c_{13}=0$, the cross-correlation function for the extra polarization (i.e., the scalar one) is rather different from the tensor one, so it is possible to use PTAs to identify the polarizations.}
The implication of the speed bounds on GWs for the generalized TeVeS theory was also considered.
The very tight speed bound drives $\tilde s_s$ to be much greater than 1, which is unnatural.
{It was also checked that the strong coupling problem does not exist in some parameter subspaces by taking into account all experimental constraints.
So the generalized TeVeS theory is excluded by the speed bounds on GWs in these parameter subspaces.}

\begin{acknowledgements}
We thank Ted Jacobson for constructive discussions.
This research was supported in part by the Major Program of the National Natural Science Foundation of China under Grant No. 11690021 and the National Natural Science Foundation of China under Grant No. 11475065.
\end{acknowledgements}

\appendix

\section{Calculating the Cross-Correlation Functions}\label{sec-app-cc}

In this appendix we present the method to calculate the cross-correlation functions for the scalar, vector, and tensor polarizations in Einstein-\ae ther theory.

\subsection{Scalar cross-correlation function}

The relative frequency shift caused by the monochromatic scalar GW is given by the first line in Eq.~\eqref{eq-fshift},
\begin{equation}\label{eq-fs-scl}
  \frac{f_e-f_r}{f_r}=\frac{(c_{14}-2c_{13})(\hat k\cdot\hat n)^2+s_s^2c_{14}(1-c_{13})}{2(1-c_{13})s_s(s_s+\hat k\cdot\hat n)}[\phi(t,0)-\phi(t-L/s_s,L\hat n)].
\end{equation}
Let the stochastic GW background be described by Eq.~\eqref{eq-sgw-scl}; thus, the timing residual is
\begin{equation}\label{eq-rt-scl}
  R(T)=\int_{-\infty}^{\infty}\frac{\ud\omega}{2\pi}\int\ud^2\hat k\left\{I_s(\hat k,\hat n)\varphi(\omega,\hat k)\frac{e^{i\omega T}-1}{i\omega}[1-e^{-i\omega L(1+\hat k\cdot \hat n/s_s)}]\right\},
\end{equation}
where
\begin{equation}\label{eq-def-sym}
  I_s(\hat k,\hat n)=\frac{(c_{14}-2c_{13})(\hat k\cdot\hat n)^2+s_s^2c_{14}(1-c_{13})}{2(1-c_{13})s_s(s_s+\hat k\cdot\hat n)}.
\end{equation}

Now, consider the correlation between two pulsars $a$ and $b$ which are at positions $\vec x_a=L_1\hat n_1$ and $\vec x_b=L_2\hat n_2$, respectively. Let $\theta=\arccos(\hat n_1\cdot\hat n_2)$ be the angular separation.
With the help of Eq.~\eqref{defvc}, the cross-correlation function between pulsars $a$ and $b$ is obtained as
\begin{equation}\label{eq-cc-ab-scl}
\begin{split}
  C_s(\theta)=&\langle R_a(T)R_b(T)\rangle\\
  =&\int_{0}^{\infty}\frac{\ud\omega}{2\pi^2}\int\ud^2\hat k\frac{|\varphi_c(\omega)|^2}{\omega^3}I_s(\hat k,\hat n_a)I_s(\hat k,\hat n_b)\mathcal{P}_s,
\end{split}
\end{equation}
where $\mathcal P_s=1-\cos\Delta_1-\cos\Delta_2+\cos(\Delta_1-\Delta_2)$ with $\Delta_j=\omega L_j(1+\hat k\cdot\hat n_j/s_s)$ ($j=1,2$).
In obtaining this result, one also averages over $T$, as implied by the ensemble average \cite{2008ApJ...685.1304L}.

If the speed $s_s$ takes the values listed in the third row of Table~\ref{tab-spds}, there will be no poles in the integrand of Eq.~\eqref{eq-cc-ab-scl}.
This is because the denominator of the integrand has a factor $(s_s+\hat k\cdot\hat n_1)(s_s+\hat k\cdot\hat n_2)$, and $\hat k\cdot\hat n_j\ge-1$,
so the denominator never vanishes.
We can approximate $\mathcal P_s=1$ whenever $\theta\ne0$, since pulsars are located at far enough distances so that the cosines in $\mathcal P_s$ oscillate fast enough and they can be ignored during the integration.
If $\theta=0$, the autocorrelation is considered by setting $\hat n_1=\hat n_2$ and $L_1=L_2$, and $\mathcal P_s\approx 2$.
In contrast, when null GWs are considered, the integrand [see Eqs.~(A36) and (A39) in Ref.~\cite{2008ApJ...685.1304L}] has at least one pole,
so $\mathcal P_s$ cannot be simply approximated as 1 or 2.

Now, one can carry out the integration by letting
\begin{gather}
  \hat n_1=(0,0,1), \label{defns-1}\\
  \hat n_2=(\sin\theta,0,\cos\theta),\label{defns-2}
\end{gather}
with the assumption that the stochastic GW background is isotropic.
Take
\begin{equation}\label{eq-set-k}
  \hat k =(\sin\theta_g\cos\phi_g,\sin\theta_g\sin\phi_g,\cos\theta_g),
\end{equation}
and so
\begin{gather}\label{deltas}
  \Delta_1=(\omega +k \cos\theta_g)L_1,\\
  \Delta_2=[\omega +k (\sin\theta_g\cos\phi_g\sin\theta+\cos\theta_g\cos\theta)]L_2.
\end{gather}
The cross correlation at $\theta\ne0$ is given by
\begin{equation}\label{eq-cc-ab-scl-ne0}
  C_s(\theta)=\int\ud\phi_g\ud\theta_g\left[I_s(\hat k,\hat n_1)I_s(\hat k,\hat n_2)\sin\theta_g\right]\int_{0}^{\infty}\ud\omega\frac{|\varphi_c(\omega)|^2}{2\pi^2\omega^3},
\end{equation}
and the autocorrelation is
\begin{equation}\label{eq-autoc-scl}
  C_s(0)=2\int\ud\phi_g\ud\theta_g\left[I_s(\hat k,\hat n_1)I_s(\hat k,\hat n_1)\sin\theta_g\right]\int_{0}^{\infty}\ud\omega\frac{|\varphi_c(\omega)|^2}{2\pi^2\omega^3}.
\end{equation}
We define the so-called normalized cross correlation $\zeta(\theta)=C_s(\theta)/C_s(0)$;
then, the frequency dependence is canceled out, so $\zeta(\theta)$ is independent of the power-law index $\alpha$.

\subsection{Vector cross-correlation function}\label{sec-vec-cors}

The relative frequency shift caused by a monochromatic vector GW is
\begin{equation}\label{eq-fs-vec}
  \frac{f_e-f_r}{f_r}=-\frac{c_{13}\hat k\cdot\hat n}{(1-c_{13})(s_v+\hat k\cdot\hat n)}[\hat n\cdot\vec v(t,0)-\hat n\cdot\vec v(t-L/s_v,L\hat n)].
\end{equation}
Now, we switch off $\mu_{\tilde 2}(\omega,\hat k)$, as the two modes $\mu_{\tilde 1}(\omega,\hat k)$ and $\mu_{\tilde 2}(\omega,\hat k)$ have an equal footing.
The timing residual caused by the stochastic vector GW background is given by
\begin{equation}\label{eq-rt-vec}
  R(T)=\int_{-\infty}^{\infty}\frac{\ud\omega}{2\pi}\int\ud^2\hat k \left\{I_v(\hat k,\hat n)\mu_{\tilde 1}(\omega,\hat k)\frac{e^{i\omega T}-1}{i\omega}[1-e^{-i\omega L(1+\hat k\cdot \hat n/s_v)}]\right\},
\end{equation}
where
\begin{equation}\label{eq-def-iv}
  I_v(\hat k,\hat n)=-\frac{c_{13}(\hat k\cdot\hat n)(\hat n\cdot\hat e_{\tilde 1})}{(1-c_{13})(s_v+\hat k\cdot\hat n)}.
\end{equation}
So the cross correlation is
\begin{equation}\label{eq-cc-vec}
  C_v(\theta)=\int_{0}^{\infty}\frac{\ud\omega}{2\pi^2}\int\ud^2\hat k\frac{|\mu^c_{\tilde 1}(\omega)|^2}{\omega^3}I_v(\hat k,\hat n_1)I_v(\hat k,\hat n_2)\mathcal P_v,
\end{equation}
where $\mathcal P_v$ can be obtained by replacing $s_s$ in $\mathcal P_s $ with $s_v$.
With $\hat k,\,\hat n_1$, and $\hat n_2$ given by Eqs.~\eqref{eq-set-k}, \eqref{defns-1} and \eqref{defns-2}, $\hat e_{\tilde{1}}$, and $\hat e_{\tilde{2}}$ are
\begin{gather}\label{eq-et1}
  \hat e_{\tilde 1}=(\cos \psi \cos \theta_g  \cos \phi_g -\sin \psi \sin \phi_g ,\cos \psi \cos \theta_g  \sin \phi_g +\sin \psi \cos \phi_g ,-\cos \psi \sin \theta_g ),\\
  \hat e_{\tilde{2}}=(-\sin \psi  \cos \theta_g  \cos \phi_g -\cos \psi  \sin \phi_g ,\cos \psi  \cos \phi_g -\sin \psi  \cos \theta_g  \sin \phi_g ,\sin \psi  \sin \theta_g ).
\end{gather}
Note that if $s_v$ takes the values in the second row in Table~\ref{tab-spds}, the integrand of Eq.~\eqref{eq-cc-vec} has no poles either.
So one approximates $\mathcal P_v$ to be 1 when $\theta\ne0$, and 2 when $\theta=0$.
The normalized cross-correlation function $\zeta(\theta)=C_v(\theta)/C_v(0)$ can thus be numerically calculated,
and it is easy to see that $\zeta(\theta)$ is independent of the power-law index $\alpha$.

\subsection{Tensor cross-correlation function}

For the tensor GWs, the relative frequency shift is
\begin{equation}\label{eq-fs-ten}
  \frac{f_e-f_r}{f_r}=\frac{s_g\hat n^j\hat n^k}{2(s_g+\hat k\cdot\hat n)}[h_{jk}^\text{TT}(t,0)-h_{jk}^\text{TT}(t-L/s_g,L\hat n)].
\end{equation}
As stated in Sec.~\ref{sec-einae-ptas}, this expression takes exactly the same form as in GR as long as $s_g=1$.
If $s_g\ne1$, this form resembles those for the massive GWs discussed in Refs.~\cite{Lee:2010cg,Lee:2014awa},
where the GW speed depends on the angular frequency through the dispersion relation.

Let us consider the cross correlation due to the plus polarization.
The timing residual of TOA is given by
\begin{equation}\label{eq-rt-tns}
  R(T)=\int_{-\infty}^{\infty}\frac{\ud\omega}{2\pi}\int\ud^2\hat k\left\{I_g(\hat k,\hat n)h_+(\omega,\hat k)\frac{e^{i\omega T}-1}{i\omega}[1-e^{-i\omega L(1+\hat k\cdot \hat n/s_g)}]\right\},
\end{equation}
where
\begin{equation}\label{eq-def-sym-g}
  I_g(\hat k,\hat n)=\frac{s_g\hat n^j\hat n^k\epsilon^+_{jk}}{2(s_g+\hat k\cdot\hat n)}.
\end{equation}
The cross correlation is thus
\begin{equation}\label{eq-cc-tns}
  C_g(\theta)=\int_{0}^{\infty}\frac{\ud\omega}{8\pi}\int\ud^2\hat k\frac{|h^+_c(\omega)|^2}{\omega^3}I_g(\hat k,\hat n_1)I_g(\hat k,\hat n_2)\mathcal P_g,
\end{equation}
in which $\mathcal P_g$ takes a similar form as $\mathcal P_s$  with $s_s$ replaced by $s_g$.
Let $s_g=1+7\times10^{-16}$, so that the integrand of Eq.~\eqref{eq-cc-tns} has no poles, and the integration can be easily done by setting $\mathcal P_g=1$ for $\theta\ne0$ and setting $\mathcal P_g=2$ for $\theta=0$. The normalized cross-correlation function $\zeta(\theta)=C_g(\theta)/C_g(0)$ can be calculated numerically.


%

\end{document}